\documentclass[12pt]{article}
\usepackage{amssymb,amsfonts,amsmath,amsthm,lineno}
\usepackage{centernot}
\usepackage[showonlyrefs]{mathtools}
\mathtoolsset{showonlyrefs=true}

\MakeRobust{\eqref}

\usepackage{hyperref}
\usepackage{enumitem}

\newcommand{\mR}{\mathbb{R}}

\newcommand{\mN}{\mathbb{N}}
\newcommand{\kk}{\mathbf{k}}
\newcommand{\pp}{\mathbf{p}}
\newcommand{\qq}{\mathbf{q}}
\newcommand{\lb}{\mathbf{l}}
\newcommand{\xx}{\mathbf{x}}

\newcommand{\vv}{\mathbf{v}}
\newcommand{\ww}{\mathbf{w}}
\newcommand{\uu}{\mathbf{u}}
\newcommand{\rr}{\mathbf{r}}

\newcommand{\pb}{\mathbf{p}}
\newcommand{\qb}{\mathbf{q}}
\newcommand{\kb}{\mathbf{k}}

\newcommand{\dsp}{\displaystyle}

\newcommand{\al}{\alpha}

\newcommand{\ga}{\gamma}
\newcommand{\vep}{\varepsilon}
\newcommand{\la}{\lambda}
\newcommand{\La}{\Lambda}
\newcommand{\p}{\partial}
\newcommand{\w}{\omega}
\newcommand{\W}{\Omega}

\newcommand{\Cc}{\mathcal{C}}
\newcommand{\Dc}{\mathcal{D}}
\newcommand{\Fc}{\mathcal{F}}
\newcommand{\Gc}{\mathcal{G}}
\newcommand{\Hc}{\mathcal{H}}
\newcommand{\Hphys}{\mathcal{H}_\mathrm{phys}}

\newcommand{\Oc}{\mathcal{O}}
\newcommand{\ov}{\overline}
\newcommand{\con}{\mathrm{const}}
\newcommand{\1}{\mathbf{1}}

\newcommand{\inc}{\mathrm{in}}

\newcommand{\ret}{\mathrm{ret}}
\newcommand{\adv}{\mathrm{adv}}
\newcommand{\as}{\mathrm{as}}
\newcommand{\reg}{\mathrm{reg}}
\newcommand{\ir}{\mathrm{ir}}
\newcommand{\fsl}[1]{{\centernot{#1}}}
\newcommand{\dW}{\dot{W}}
\newcommand{\dV}{\dot{V}}
\newcommand{\dU}{\dot{U}}
\newcommand{\ti}{\widetilde}

\DeclareMathOperator{\id}{id}
\DeclareMathOperator{\Ker}{Ker}

\DeclareMathOperator{\sgn}{sgn}

\DeclareMathOperator*{\slim}{s-lim}
\DeclareMathOperator*{\wlim}{w-lim}
\DeclareMathOperator{\Tr}{Tr}

\newtheorem{thm}{Theorem}

\newtheorem{lem}[thm]{Lemma}
\newtheorem{col}[thm]{Corollary}

\usepackage[a4paper, total={14cm, 24cm}]{geometry}

\title{Undressing the electron}
\author{Andrzej Herdegen\thanks{e-mail: herdegen@th.if.uj.edu.pl}\\
{\it Institute of Physics, Jagiellonian University,}\\
{\it ul.\,S.\,{\L}ojasiewicza 11, 30-348  Krak\'{o}w, Poland}}\date{}

\begin{document}

\maketitle

\begin{abstract}
The extended algebra of the free electromagnetic fields, including infrared singular fields, and the almost radial gauge, both introduced earlier, are postulated for the construction of the quantum electrodynamics in a Hilbert space (no indefinite metric). Both the Dirac and electromagnetic fields are constructed up to the first order (based on the incoming fields) as operators in the Hilbert space, and shown to have physically well interpretable asymptotic behavior in far past and spacelike separations. The Dirac field tends in far past to the free incoming field, carrying its own Coulomb field, but with no `soft photon dressing'. The spacelike asymptotic limit of the electromagnetic field yields a conserved operator field, which is a sum of contributions of the incoming Coulomb field, and of the low energy limit of the incoming free electromagnetic field. This should agree with the operator field similarly constructed with the use of outgoing fields, which then relates these past and future characteristics. Higher orders are expected not to change this picture, but their construction needs a treatment of the UV question, which has not been undertaken and remains a problem for further investigation.

\end{abstract}

\maketitle

\section{Introduction}

One of the fundamental paradigms of quantum field theory (QFT) is the relativistic locality, which in its actual form used in this field may be characterized by the following statements.\\
1/ Basic observables of the theory represent quantities which may be measured in bounded spacetime regions. In the algebraic formulation of the QFT this structure is formalized in the form of a net of C*-algebras, while both in the Wightman axiomatic and in the practical formulation of the perturbational approach, these objects are operator-valued distributions smeared with functions of compact support, and representing measurable quantities like electromagnetic field and electric current density vector. \\
2/ If $A_1$ and $A_2$ are local observables ascribed in this sense to two spatially separated regions $\Oc_1$ and $\Oc_2$, respectively (all vectors $x_2-x_1$, with $x_i\in\Oc_i$, are space-like), then $[A_1, A_2]=0$.\\
3/ If an operator commutes  with all local observables, then it is proportional to the identity operator.

The latter condition defines what is called a superselection sector of the theory, in which local observables are represented irreducibly. Different superselection sectors describe physical settings which are different in some global sense. In the algebraic approach it is claimed that local observables are all one needs to physically characterize a theory (see \cite{haa06}). However, for the construction of a particular theory it also proved indispensable to employ fields subject to gauge transformations, such as the electromagnetic potential, and the Dirac field, which should create electric charge and interpolate between different superselection sectors. In recent decades this has been recognized in the algebraic approach as well (see a recent summary \cite{rej16}). As the locality paradigm, beside its physical role, proved crucial for the techniques employed for solving ultra-violet problems of QFT, these `unphysical' fields have been included into its range (with the replacement of commutator by anticommutator for two fermion fields). However, in theories with constraints and long-range interactions, such as electrodynamics, this technical measure contradicts physics: a `local Dirac field' applied to a neutral state cannot change electromagnetic field at spatial infinity, so the charge calculated by Gauss' law does not change. Therefore, the local construction may only be a first step for the construction of physical states, which do not belong to the original state space of the construction (as in the analysis by Steinmann \cite{sta00}).

Let us now look more closely at some of the consequences of the above structure for quantum electrodynamics (QED), in which the Maxwell equations hold,\footnote{Spacetime indices will be denoted by $a,b,c$, etc. We assume that an origin in Minkowski space $M$ has been chosen, and then points are represented by vectors $x$, $y$ etc. Moreover, we choose a unit, timelike, future-pointing vector $t$, to define the time axis, and then $x^0=x\cdot t$, $\xx$ is the three-space part of $x$,  and  $|x|=(|x^0|^2+|\xx|^2)^\frac{1}{2}$.\\ We use the Gauss units; while Heaviside's `rationalization' removes $4\pi$ from the Maxwell equations, it makes the formalism to be used later inconveniently plagued by such factors elsewhere.\label{int_conv}}
\begin{equation}\label{int_maxwell}
 \p_{[a}F_{bc]}(x)=0\,,\qquad \p^aF_{ab}(x)=4\pi j_b(x)\,,
\end{equation}
in the operator-distributional sense. The most natural selection condition for the representations of physical interest is to demand that the decay of the electromagnetic field at spatial infinity is of the Coulomb field rate, i.e.\ for spacelike $x$, and $R$ tending to infinity, the field $R^2F_{ab}(Rx)$ has a well defined, distributional limit. More precisely,
let $\Hphys$ be a dense subspace of (physical) vector states in the Hilbert space $\Hc$ of the superselection sector, contained in the domains of the smeared fields   $F_{ab}(\varphi)$, and stable under the action of all local observables (including the unbounded field $F_{ab}(\varphi)$). Then we assume that for each $\varphi$ of compact, spacelike support, there exists a self-adjoint operator $F^0_{ab}(\varphi)$, with $\Hphys$ contained in the domain, such that for all $\Phi\in\Hphys$ there exist weak limits
\begin{equation}\label{int_tail}
 \wlim_{R\to\infty}R^{-2}F_{ab}(\varphi_R)\Phi
 =F^0_{ab}(\varphi)\Phi\,,
\end{equation}
where $\varphi_R(x)=\varphi(x/R)$. However, for large $R$ the support of $\varphi_R$ becomes spacelike with respect to any chosen compact region. Therefore, $F^0_{ab}(\varphi)$ commutes on $\Hphys$ with all local observables, so by irreducibility is, in fact, a numerical distribution in $x\cdot x<0$,  homogeneous of degree $-2$.\footnote{The well-known analysis by Buchholz \cite{buch86} derives \eqref{int_tail}, here postulated, from an assumption on the limit of the mean value, and boundedness of fluctuations, in \emph{one} chosen state.} This shows that a physical representation of the assumed type is characterized by a classical flux of the electromagnetic field at spatial infinity. Two such sectors with different fluxes carry inequivalent representations of observables.

For the derivation of further consequences one needs to represent solutions of \eqref{int_maxwell} in the form suitable for scattering processes. Assume that the convolution\footnote{For instance, in the sense described by Vladimirov \cite{vla02}.} $D^\ret\ast j_b$ exists as an operator-valued distribution, so that one can define the retarded electromagnetic field by
\begin{equation}\label{int_ret}
 F^\ret_{ab}(\varphi)= 4\pi[D^\ret\ast (\p_aj_b-\p_bj_a)\,](\varphi)
 =4\pi\int\limits_M [D^\adv\ast\varphi\,](x)(\p_aj_b-\p_bj_a)(x)dx\,,
\end{equation}
where $dx$ is the standard Lebesgue measure on the Minkowski space $M$
(note that the existence assumption involves only long-range behavior of $j_b(x)$, and not its local properties, as $D^\adv\ast\varphi$ is a smooth function). Then the decomposition of $F_{ab}$ suitable for the discussion of incoming asymptotics is
\begin{equation}\label{int_retin}
 F_{ab}=F^\ret_{ab}+F^\inc_{ab}\,,
\end{equation}
where $F^\inc_{ab}$ satisfies the homogeneous Maxwell equations and describes the incoming radiation field.

For the discussion of the spacelike, or the past, asymptotics based on the above splitting, one needs further assumptions on incoming charged particles. As mentioned above, creation of a charged particle is a nonlocal operation, which has the consequence that one-particle states do not correspond to a discrete mass value in the energy-momentum spectrum (infraparticle problem). This problem has been shown in \cite{buch86} to follow from the above described superselection structure and Gauss' law alone. Charged particles differ from neutral ones. However, if the particle interpretation of QED should be possible at all, and the charge current is carried by massive particles only, then at least the following assumption  seems reasonable. For $\chi(x)$ a test function with support outside the future light cone $V_+$, $\chi_R(x)=\chi(x/R)$, and $\Phi_i\in\Hphys$, one should have
\begin{equation}\label{int_ascur}
  \lim_{R\to\infty}R^{-1}(\Phi_1,j_b(\chi_R)\Phi_2)
 =(\Phi_1,j^{\inc,0}_b(\chi)\Phi_2)\,,
\end{equation}
where $j^{\inc,0}_b(x)$ is an operator-valued distribution, homogeneous of degree $-3$, with support inside the past light cone $V_-$. This incorporates minimal expectations for incoming charges moving freely. By a slight strengthening of this, one should expect that for support of $\varphi$ in large separation (timelike or spacelike) from $V_+$, the current $j_a$ may be replaced by $j^{\inc,0}_a$ in \eqref{int_ret}. Therefore, both the past and  the spacelike asymptotics of $F^\ret_{ab}$ are determined by $j^{\inc,0}_a$.
In the spacelike case, one can thus expect
\begin{equation}\label{int_retail}
\begin{gathered}
 \lim_{R\to\infty}R^{-2}(\Phi_1,F^\ret_{ab}(\varphi_R)\Phi_2)
 =(\Phi_1,F^{\ret,0}_{ab}(\varphi)\Phi_2)\,, \\
 F^{\ret,0}_{ab}=4\pi D^\ret\ast (\p_aj^{\inc,0}_b-\p_bj^{\inc,0}_a)\,.
\end{gathered}
\end{equation}

The spacelike asymptotic structure described above has now consequences which form the core of our discussion. It follows from relations \eqref{int_tail} and \eqref{int_retail} that in the assumed context also the radiation field $F^\inc_{ab}$ must have the spacelike asymptotics of the form
\begin{equation}\label{int_intail}
 \lim_{R\to\infty}R^{-2}(\Phi_1,F^\inc_{ab}(\varphi_R)\Phi_2)
 =(\Phi_1,F^{\inc,0}_{ab}(\varphi)\Phi_2)\,,
\end{equation}
where $F^{\inc,0}_{ab}(x)$ is an operator-valued distribution in $x\cdot x<0$, homogeneous of degree $-2$. In consequence,
\begin{equation}\label{int_sumtail}
 F^{\ret,0}_{ab}(x)+F^{\inc,0}_{ab}(x)=F^0_{ab}(x)\id\,.
\end{equation}
This relation shows that these fields $F^{\ret,0}_{ab}$ and $F^{\inc,0}_{ab}$ cannot be statistically independent, as the sum in \eqref{int_sumtail} is classical and fixed. Taking expectation values of this relation in various states $\Phi$ of this representation, one finds that in the given representation, characterized by the numerical distribution $F^0_{ab}(x)$, each configuration of incoming charges must be  accompanied by a `cloud' of radiation so fine-tuned that the spacelike tail of the electromagnetic field remains unchanged. Therefore, the field $F^{\inc,0}_{ab}(x)$ in \eqref{int_sumtail} has only a `slave' status with respect to $F^{\ret,0}_{ab}(x)$, with its only function of keeping $F^0_{ab}(x)$ fixed.\footnote{This conclusion converges with the results of more extensive algebraic analyses, see \cite{buch82}, and \cite{haa06}, Sections VI.2 and VI.3.}

The picture emerging from the above scheme may suggest that the behavior of the electromagnetic field in spacelike infinity, if only consistent with the Gauss law, is largely a matter of convention. Another scheme, recently put forward, is based on explicit acceptance of this view. One considers `infravacuum' representations, in which electromagnetic field has infinite fluctuations in spacelike infinity, which prevents the existence of the limit \eqref{int_tail}. It is further claimed that this behavior is not in conflict with experiments, as they can be performed only inside future light  cones.\footnote{The idea of restricting observable physics to a light cone has been put forward in \cite{buch14}; see also \cite{duch23}.}

In this paper we continue our explorations oppositely directed, towards full inclusion of the long range variables into the quantum observable realm. We believe that the long-range structure and the existence of constraints in electrodynamics justify such expectations.  The plan of the article is as follows. In Section~\ref{main} we explain in heuristic terms the ideas behind our constructions. Sections~\ref{direlm} and \ref{ar} summarize, for the convenience of the reader, the representation of the basic variables of the theory (a new addition is the explicit Fock form of the representation of the free electromagnetic field, implicit in earlier works). In Section \ref{inter} we construct the first order fields. After describing in Section \ref{transl} the translation automorphism of fields, we investigate various asym\-ptotic limits of fields in Sections \ref{asd} -- \ref{null}. Sections 5, 7, 8, and 9 constitute the advance of our programme. Physical discussion of the results and conclusions are gathered in the closing Section \ref{diss}. Appendices contain some supplementary material.

\section{Main ideas}\label{main}

The purpose of this article is to give preliminary steps for the perturbative construction of the Maxwell-Dirac quantum electrodynamics in a physical gauge in a Hilbert space (no indefinite metric), in which:
\begin{itemize}[topsep=3pt, partopsep=1pt, itemsep=0pt, leftmargin=18pt]
\item[(i)] the incoming infrared-singular electromagnetic fields have full quantum status,
\item[(ii)] the interacting fields have well-defined incoming asymptotics.
\end{itemize}

These steps include the construction of the Dirac, and the electromagnetic field, in the first order of perturbation. This will allow us to make some conjectures on the structure. Further development necessitates treatment of UV problems in our scheme, and we leave this for future investigation.

The construction has two key elements, each addressing one of the two above points.
\begin{itemize}[topsep=2pt, partopsep=1pt, itemsep=0pt, leftmargin=18pt]
\item[(i)] The construction of the free electromagnetic fields, including those with nontrivial infrared tails (``infrared-singular'' fields), uses the extension of the local algebra defined in \cite{her98}, and its positive energy representations constructed in the same article.\footnote{A recent summary of this construction and its further development is discussed  in \cite{her17}.} This algebra contains a nonlocal element constituted by the infrared tails, which has the consequence that it admits no vacuum state. Instead, there do exist states which have arbitrarily small energy content. This residual energy cannot be wholly ascribed to photons---the particle interpretation of the electromagnetic field is not complete. The use of this construction makes it possible to free the incoming free field from its `slave' dependence described in Introduction.

\item[(ii)] The long-range character of the electromagnetic interaction implies, as is well known, difficulties in time asymptotics of scattering.\footnote{The Dollard technique, which is very fruitful in nonrelativistic potential quantum mechanics \cite{der97}, when applied in QED allows to construct the scattering operator, but does not solve the problem of the construction of interacting fields in a physical gauge. See recent constructions developed in \cite{duc19}.} It has been shown in \cite{her21} that in classical electrodynamics of the Dirac field in the external electromagnetic field (with plausible extension to the full interacting theory) the Dirac field does tend asymptotically to its free versions, if an appropriate gauge is chosen. The crucial constituting property of this gauge is that $x\cdot A(x)$ vanishes sufficiently fast for $x$ tending to past and future infinity inside the light cone. A quantum version of such gauge, for a free field in the algebra mentioned above in (i), has been constructed recently \cite{her22}, and given the name of `almost radial gauge'.
\end{itemize}

The heuristic rationale behind the gauge condition described above, may be understood as follows. Asymptotically in time the electromagnetic current is carried by charged particles moving freely towards/away from the region of interaction around $x=0$, thus $j(x)$ becomes proportional to $x$. With the above gauge condition the action integral of interaction $\int j\cdot A\, dx$ becomes finite.

We now sketch further steps in loose terms. The solution of the Maxwell-Dirac system consisting of the Maxwell equations \eqref{int_maxwell}, with the current $j_b(x)=e\bar{\psi}(x)\gamma_b\psi(x)$, and the Dirac equation
\begin{equation}\label{int_dirac}
 [\ga^a(i\p_a-eA_b(x))-m]\psi(x)=0\,,
\end{equation}
where $e<0$ is the electron charge, is selected by rewriting them in the form of integral equations
\begin{align}
 \psi(x)
 &=\psi^\inc(x)
 -e\int_M S^\ret(x-y)\ga^bA_b(y)\psi(y)\, dy\,,\label{int_dir}\\
 A_a(x)&=A^\inc_a(x)
 +4\pi e\int_M \Dc^\ret_{ab}(x,y)\bar{\psi}(y)\ga^b\psi(y)\,dy\,,\label{int_pot}
 \end{align}
where
\begin{equation}
 S^\ret(x)=(2\pi)^{-4}\int\frac{(\fsl{k}+m)e^{-ik\cdot x}}{m^2-k^2-i\sgn(k^0)\,0}\,dk
\end{equation}
is the retarded fundamental solution of the Dirac equa\-tion, which satisfies  \mbox{$[i\ga\cdot \p -m]S^\ret(x)=-\delta(x)$}, and we shall later comment on the choice of\linebreak $\Dc^\ret_{ab}(x,y)$.\footnote{The renormalization of products poses no problem in the lowest order to be considered.} We shall show that up to the first order, in contrast to the usual formulations, one has $\psi(x)\to \psi^\inc(x)$ in far past in appropriate sense. We conjecture that the higher orders will not change this limit. Also, we shall argue that the spacelike and past asymptotics of the electromagnetic field is determined by the solution for $A(x)$ up to the first order, and is a sum of $A^\inc(x)$ and the  Coulomb (`bare') field of asymptotic particles.

We end this section by fixing some further conventions and notation. The future directed hyperboloid and the light cone in Minkowski space will be denoted $H_m=\{p:p^2=m^2, p^0>0\}$ and $C_+=\{l:l^2=0, l^0>0\}$, respectively. Intrinsic differentiation operators in these hypersurfaces are
\begin{equation}
 M_{ab}=p_a\frac{\p}{\p p^b}-p_b\frac{\p}{\p p^a}\,,\quad
 L_{ab}=l_a\frac{\p}{\p l^b}-l_b\frac{\p}{\p l^a}\,.
\end{equation}
For functions on $C_+$ with the homogeneity property $f(\la l)=\la^{-2}f(l)$, $\la>0$, the integral on the set of null directions defined by
\begin{equation}
 \int f(l)\,d^2l=\int f(1,\lb)\,d\W(\lb)\,,
\end{equation}
where $d\W(\lb)$ is the spherical angle measure on the unit sphere, is Lorentz-invariant (does not depend on the reference system). Consequently,
\begin{equation}\label{int_Lint}
 \int L_{ab}f(l)\,d^2l=0\,,\quad \int \p\cdot V(l)\,d^2l=0\,,
\end{equation}
where in the second integral $V(l)$ is a homogeneous function of degree $-1$,  such that $l\cdot V(l)=0$, and extended for the sake of differentiation to a neighborhood of the cone with preservation of these two properties. For the latter property we refer the reader to \cite {her98},  \cite{her17}. Further information on homogeneous functions on $C_+$ may be found in Appendix \ref{cone}.

Let $\chi(x)$ be a Minkowski space field, and $W(s,l)$ be a function of $s\in\mR$ and $l\in C_+$. Then we denote
\begin{equation}
\begin{aligned}
 \hat{\chi}(p)&=\frac{1}{2\pi}\int_M e^{ip\cdot x}\chi(x)\,dx\,,\\
 \ti{W}(\w,l)&=\frac{1}{2\pi}\int_\mR e^{i\w s} W(s,l)\,ds\,.
\end{aligned}
\end{equation}
Moreover, we write $\dW(s,l)=\p_sW(s,l)$.

In what follows, whenever the range of integration is suppressed, the whole domain of the  measure is meant.

\section{Algebra and representation of incoming\\ free fields}\label{direlm}

The Hilbert space of the model is the tensor product $\Fc=\Fc_D\otimes\Fc_R$, where $\Fc_D$ carries the standard Fock representation of the free Dirac field, while $\Fc_R$ is the representation space of the free incoming electromagnetic field. To fix our notation we first write down the basic formulae of the spinor field, and then review our construction of the electromagnetic field and the almost radial gauge potential.

\subsection{Dirac field}\label{dir}

For $p\in H_m$, the mass hyperboloid, we denote the projection operators $P(\vep p)$, $\vep=\pm$, on the bispinor space:
\begin{equation}\label{dir_P}
 P(\vep p)=\frac{1}{2m}(m+\vep\ga\cdot p)\,,\ P(+p)+P(-p)=1\,,\  P(\vep_1p)P(\vep_2 p)=\delta_{\vep_1,\vep_2}P(\vep_1p)\,.
\end{equation}
If we write the Dirac in-field in the form
\begin{equation}
 \psi^\inc(x)=(2\pi)^{-\frac{3}{2}}
 \int\sum_{\vep=\pm}
 e^{-\vep ix\cdot p}c_\vep(p)\,d\mu_m(p)\,,\quad
 c_\vep(p)=P(\vep p)c(p)\,,
\end{equation}
then the anticommutation relations have two equivalent forms:
\begin{align}
 [\psi^\inc(x),\bar{\psi}^\inc(y)]_+&=-iS(x-y)\\
 &=(2\pi)^{-3}\int\delta(k^2-m^2)\sgn(k^0)
 (\fsl{k}+m)e^{-ik\cdot x}\,dk\,,\\
 [c(p),\bar{c}(k)]_+&=\frac{\ga\cdot p}{m}\,\delta_m(p,k)\,,
\end{align}
where (with $p$ on the mass shell)
\begin{equation}\label{dir_md}
 d\mu_m(p)=\frac{m}{p^0}\,d^3\pp\,,\quad \delta_m(p,k)=\frac{p^0}{m}\delta(\pp-\kk)\,.
\end{equation}
The operators $c_+(p)$ and $c^*_-(p)$ are the annihilation operators of the electron and the positron, respectively, and their adjoints are the respective creation operators. The subspace of $\Fc_D$ consisting of all finite particle vectors with momentum profiles of the Schwartz type on $H_m$ will be denoted by $\Fc_D^S$. The operators $c(p)$ and $\bar{c}(p)$, when smeared with Schwartz functions, act intrinsically in $\Fc_D^S$.

The Fourier transform of $j^\inc(x)=e:\bar{\psi}^\inc(x)\ga\psi^\inc(x):$, the current of the in-field, takes the form
\begin{equation}\label{dir_curr}
 \hat{j}^\inc(k)=e\int \sum_{\vep_1,\vep_2}
 :\bar{c}_{\vep_1}(p)\ga c_{\vep_2}(q):\delta(k+\vep_1p-\vep_2q)\,d\mu_m(p)\,d\mu_m(q)\,,
\end{equation}
and has a well-defined restriction to the light cone, $k=\w l$, $\w\in\mR$, $l\in C_+$:
\begin{equation}
 \hat{j}^\inc(\w l)=\delta(\w)
 \int\rho_D(p)
 \frac{p}{p\cdot l}\,d\mu_m(p)\,.
\end{equation}
where
\begin{equation}\label{dir_dench}
 \rho_D(p)=e\sum_{\vep=\pm}\vep :\bar{c}_\vep(p)c_\vep(p):=em^{-1}:\bar{c}(p)p\cdot\ga c(p):
\end{equation}
is the operator of the momentum-density of charge.

\subsection{Electromagnetic field}\label{elm}

The extended electromagnetic algebra is built of elements $A(J)$, which symbolize potential $A(x)$ smeared with conserved, charge-free test currents $J(x)$. These elements are gauge-invariant; thus they represent electromagnetic fields. For the algebra of incoming fields the admitted currents may extend in past timelike directions, with the timelike tails of the form $J(x)\sim x\sigma(x)$, with $\sigma(x)$ homogeneous of degree $-4$, with past timelike support.\footnote{For the outgoing field, future timelike tails would be admitted.} The standard commutation symplectic form is then not absolutely integrable, but may be extended in the following way:
\begin{equation}\label{elm_com}
 [A(J_1),A(J_2)]=\frac{i}{2}\int [J_1\cdot A_2-J_2\cdot A_1](x)\,dx\,,
\end{equation}
where $A_i(x)=4\pi\int D(x-y)J_i(y)\,dy$, with $D(x)$ the Pauli-Jordan function. This algebra describes free electromagnetic fields, including these with typical infrared tails of the Coulomb decay rate, as in the case of radiation fields produced by scattered charges. If for each current $J$ we define ($s\in\mR$, $l\in C_+$)
\begin{equation}\label{elm_VJ}
 V(s,l)=\int\delta(s-x\cdot l)J(x)\,dx\,,\quad \text{so}\quad
 \ti{V}(\w,l)=\hat{J}(\w l)\,,
\end{equation}
then for the incoming test currents of the form mentioned above the functions vanish for $s\to\infty$, and have limits $V(-\infty,l)$ for $s\to-\infty$. If we now use the representation of the Pauli-Jordan function \eqref{ar_PJ} given below, then
\begin{equation}
 \tfrac{1}{2}\int [J_1\cdot A_2-J_2\cdot A_1](x)\,dx
 =\{V_1,V_2\}\,,
\end{equation}
where
\begin{equation}\label{elm_sform}
\begin{aligned}
 \{V_1,V_2\}&=\frac{1}{4\pi}\int [\,\dV_1\cdot V_2-V_1\cdot\dV_2\,](s,l)\,ds\,d^2l\\
 &=\frac{1}{4\pi}\int\dV_1(s,l)\sgn(s-\tau)\dV_2(\tau,l)\,ds\,d\tau\, d^2l\\
 &=i\int\frac{\ov{\ti{\dV}_1}\cdot\ti{\dV}_2(\w,l)}{\w}\,d\w d^2l\,,
\end{aligned}
\end{equation}
where the second integral representation uses the fact that both $V_i(s,l)$ vanish in $s=+\infty$, and the third integral is in the principal value sense.
Introducing notation $A(J)=\{V,V^\mathrm{q}\}$ we represent the field $A$ by a quantum variable $V^q(s,l)$, and then the commutation relation \eqref{elm_com} takes the form
\begin{equation}\label{elm_algV}
 \big[\{V_1,V^\mathrm{q}\},\{V_2,V^\mathrm{q}\}\big]=i\{V_1,V_2\}\,.
\end{equation}
The functions $\ti{\dV}(\w,l)$ have the following properties
\begin{gather}
 l\cdot \ti{\dV}(\w,l)=0\,,\qquad
 \ti{\dV}(\la^{-1}\w,\la l)=\la^{-1}\ti{\dV}(\w,l)\quad
 \text{for all}\quad \la>0\,,\label{elm_propV}\\
 L\wedge \ti{\dV}(0,l)=0\,.\label{elm_propV0}
\end{gather}
Moreover, for the sake of the present analysis we restrict them to be of Schwartz type. Condition \eqref{elm_propV0}, which is a consequence of the asymptotic behavior of $J$, has an equivalent form: there exists a real, homogeneous function $\Phi(l)$ such that
\begin{equation}
 \ti{\dV}(0,l)=\p\Phi(l)\,.
 \end{equation}
 Note that both $\ti{\dV}(0,l)$ and $\p\Phi(l)$ are real functions (as $\dV(s,l)$ are real).

Translation-covariant, positive energy representations of this algebra (in the $C^*$ Weyl form) were constructed in \cite{her98}. The idea, roughly, may be described as follows. Consider first the usual algebra of IR-regular fields, and take all its coherent state representations, characterized by  spacelike tails of additional classical fields of the type created in scattering of charges. Choose a Gaussian measure on the space of these tails and form the direct integral of these coherent state representations. If the covariance operator of this measure satisfies certain regularity conditions, then on the Hilbert space of this construction the extended electromagnetic algebra is naturally represented in a regular, translation-covariant  (with positive energy), and irreducible way.

Representation thus constructed has no vacuum vector, nor any other canonical choice of a state. However, there does exist a class of states $\w_h$ on the algebra (with $h$ going over a class of functions introduced before \eqref{elm_Vh} below), each of which defines a unitary equivalent representation of a Fock type, based on this state.\footnote{In the original representation constructed in Section 6 in \cite{her98}, this state is represented by a particular vector (6.11) in this reference, with $\chi([f])=1$, $\psi=\W$---the Fock `vacuum' of the space $\hat{\Hc}_{[0]}$.
The equivalence of representations is based on Theorem 6.4 and Proposition 6.5. We warn the reader that there is a misprint in Eq.\ (6.16), there should be the direct product, instead of direct sum symbol, between the representations $\pi_\mu$ and $\hat{\pi}_{[0]}$. Also, in the exposed formula after (6.23) one should have $B^{-1/2}$ instead of $B^{1/2}$.} The state $\w_h$ is represented by the Fock `vacuum' in this representation. However, we stress that this state is not translationally invariant, and we discuss its energy-momentum content in Appendix \ref{omega}. This Fock form of the representation is most useful for calculations and we briefly sketch the construction.

The Hilbert space of this representation is constructed as follows. First, let $F(\w,l)$ denote a complex vector function on $\w\in[0,\infty)$, $l\in C_+$, orthogonal to~$l$, $l\cdot F(\w,l)=0$, and with the property
$F(\la^{-1}\w,\la l)=F(\w,l)$ for all $\la>0$. Then $\Hc_\reg$ is the Hilbert space of such functions square integrable with respect to the scalar product
\begin{equation}\label{elm_scalF}
 (F_1,F_2)_\reg=-\int\limits_{[0,\infty)\times C_+} \ov{F_1}\cdot F_2(\w,l)\,\w\,d\w\,d^2l\,.
\end{equation}
More precisely, each $F(\w,l)$ symbolizes an equivalence class of functions differing by additions of functions proportional to $l$, but the use of the above simplified notation does not lead to inconsistencies.
Next, let $f(l)$ denote a complex function on $C_+$, homogeneous of degree $-1$,  orthogonal to $l$, $l\cdot f(l)=0$, and consider the Hilbert space $\Hc_0$ of such functions (again, more precisely equivalence classes with respect to addition of terms proportional to $l$) square integrable with respect to the product
\begin{equation}\label{elm_scalf}
 (f_1,f_2)_0=-\int\limits_{C_+} \ov{f_1}\cdot f_2(l)\,d^2l\,.
\end{equation}
Functions of the form $f(l)=\p\phi(l)$, where $\phi(l)$ is a smooth and homogeneous of degree $0$ complex function, form a subspace of $\Hc_0$. We denote by $\Hc_\ir$ the closure of this subspace in $\Hc_0$, and by $P_\ir$ the orthogonal projection operator from $\Hc_0$ onto $\Hc_\ir$; the explicit form of this operator is derived in Appendix \ref{cone}. For $f_1,f_2\in\Hc_\ir$ we shall write $(f_1,f_2)_\ir=(f_1,f_2)_0$.
We form the direct sum Hilbert space $\Hc_R=\Hc_\ir\oplus\Hc_\reg$, and the Hilbert space of the representation is now the Fock space $\Fc_R$ based on the `one-excitation' space $\Hc_R$. As well known, another way of forming this space is $\Fc_R=\Fc_\mathrm{ir}\otimes\Fc_\reg$, where $\Fc_\ir$ and $\Fc_\reg$ are Fock spaces based on $\Hc_\ir$ and $\Hc_\reg$, respectively. The pairs
$a_\ir(f), a_\ir^*(f)$ and $a_\reg(F), a_\reg^*(F)$ are creation/annihilation operators in $\Fc_\ir$ and $\Fc_\reg$, respectively, satisfying standard commutation relations
\begin{equation}
 \big[a_\ir(f_1),a_\ir^*(f_2)\big]=(f_1,f_2)_\ir\,,\qquad \big[a_\reg(F_1),a_\reg^*(F_2)\big]=(F_1,F_2)_\reg\,.
\end{equation}
In what follows, the subscripts `$\ir$' and `$\reg$' at scalar products will be often omitted, if it does not lead to confusion.
The operators
\begin{equation}\label{elm_aaa}
 a^\#(f,F)=a^\#_\ir(f)\otimes \1+\1\otimes a^\#_\reg(F)
\end{equation}
are creation/annihilation operators in $\Fc_R$.\footnote{To simplify notation we shall also write $a_\ir^\#(f)\otimes\1\equiv a_\ir^\#(f)$ and
$\1\otimes a_\reg^\#(F)\equiv a_\reg^\#(F)$.}
We denote by $\Fc^S_R\subset \Fc_R$ the subspace formed of all finite-excitation vectors, with wave functions smooth on their domains $(C_+)^{\times k}\times([0,\infty)\times C_+)^{\times n}$, and vanishing fast together with all their derivatives in infinity of $\w\in[0,\infty)$. Creation/annihilation operators with $(f,F)$ restricted to one-excitation subspace of $\Fc^S_R$ act intrinsically on $\Fc^S_R$.

For practical purposes, it is convenient to reinterpret the space $\Hc_\ir$ in a unitary equivalent way, by the transformation $\p\phi(l)\mapsto [\phi(l)]$, where $[\phi(l)]$ denotes the equivalence class with respect to the addition of a constant, an element of the Hilbert space $\Hc_{\p^2}$---completion with respect to the product
\begin{equation}
 ([\phi_1],[\phi_2])_{\p^2}=\int\ov{\phi_1}(l)\p^2\phi_2(l)\,d^2l
 =-\int\p\ov{\phi_1}(l)\cdot \p\phi_2(l)\,d^2l\,.
\end{equation}
When not leading to uncertainty, the brackets at $[\phi]$ will be omitted.
Operators acting in $\Hc_\ir$ are easier defined by their versions acting in $\Hc_{\p^2}$, the relation being
\begin{equation}
 C\to C_{\p^2}\,,\qquad C\p\phi=\p(C_{\p^2}\phi)\,.
\end{equation}

Let now $\Hc_\ir^r$ be the real Hilbert space of real functions in $\Hc_\ir$, and denote by $\Cc^r_\ir\subset\Hc^r_\ir$ the subspace of smooth functions. We choose an operator $B$ (and the corresponding operator $B_{\p^2}$) with the following properties:
\begin{equation}\label{elm_B}
\begin{gathered}
 B:\Hc_\ir^r\to\Hc_\ir^r\,,\quad \Ker B=\{0\}\,,\quad B>0\,,\quad
 \Tr B<\infty\,,\quad B^\frac{1}{2} \Cc^r_\ir=\Cc^r_\ir\,.
 \end{gathered}
 \end{equation}
Operator $B$ with these properties is the covariance operator of the Gaussian measure used for the construction of the representation formed by integration of the coherent state representations, as briefly described above.

A unitarily equivalent Fock representation is now obtained as follows.
Choose a real Schwartz function $h(s,l)$ on $\mR\times C_+$, homogeneous of degree $-1$, and such that its Fourier transform $\tilde{h}(\w,l)$ satisfies $\tilde{h}(0,l)=1$. For each Schwartz function $\tilde{\dV}(\w,l)$ with properties \eqref{elm_propV} and \eqref{elm_propV0} denote
\begin{align}
 \ti{V}_h(\w,l)&=i\w^{-1}\big[\ti{\dV}(\w,l)-\ti{\dV}(0,l)\ti{h}(\w,l)\big]\,,
 \label{elm_Vh}\\
 p(\dV)(l)&=\ti{\dV}(0,l)\,,\label{elm_pV}\\
 r_h(\dV)(l)&=P_\ir\,R_h(\dV)(l)\,,\label{elm_rV}\\ R_h(\dV)(l)&=-i\int_{\mR}\ov{\ti{h}}(\w,l)\ti{\dV}(\w,l)\frac{d\w}{\w}
 =-\int_{\mR}\ov{\ti{h}}(\w,l)\ti{V}_h(\w,l)\,d\w\,,\label{elm_RV}\\
 j_h(\dV)&=\tfrac{1}{2}B^{-\frac{1}{2}}p(\dV)+iB^\frac{1}{2}r_h(\dV)\,,\label{elm_jV}
\end{align}
where $P_\ir$ in \eqref{elm_rV} is the projection defined above, and the first integral in \eqref{elm_RV} is in the principal value sense; note that $p(\dV)$ and $r_h(\dV)$ are real functions. Then the  formula\footnote{When $\ti{V}_h(\w,l)$ is an argument of $a^\#(\ti{V}_h)$, it is understood that its restriction to $\w\geq0$ is meant.}
\begin{equation}\label{elm_repr}
 \pi_h(\{V,V^\mathrm{q}\})
 =a(j_h(\dV), \ti{V}_h)+a^*(j_h(\dV), \ti{V}_h)
\end{equation}
defines a representation of relations \eqref{elm_algV} acting in $\Fc^S_R$. The von Neumann algebra of its exponentiation acts irreducibly in $\Fc_R$, and is unitarily equivalent to the representation briefly characterized above. Indeed, one easily finds that
\begin{equation}
 \big[\pi_h(\{V_1,V_1^\mathrm{q}\}), \pi_h(\{V_2,V_2^\mathrm{q}\})\big]
 =i\{V_1,V_2\}\,,
\end{equation}
and if we denote by $\Omega_h$ the Fock `ground state' of $\Fc_R$ then
\begin{equation}
 \big(\Omega_h, [\pi_h(\{V,V^\mathrm{q}\})]^2 \Omega_h\big)
 =\|\ti{V}_h(\w,l)\|^2_\reg + \tfrac{1}{4}\|B^{-\frac{1}{2}}p(\dV)\|^2_\ir
 +\|B^\frac{1}{2}r_h(\dV)\|^2_\ir\,.
\end{equation}
The latter formula defines the quasi-free state determined by Theorem 6.4 and Proposition 6.5 in \cite {her98}; see also Section 4.4 in \cite{her08}. Moreover, this equivalence also guarantees that the representations defined as above, with different functions $h$, are unitarily equivalent.\footnote{More precisely, one can show that they are related by an implementable Bogoliubov transformation, but we do not discuss the details here.}

As anticipated at the beginning of the present section, the Hilbert space of the theory is now $\Fc=\Fc_D\otimes\Fc_R$. The sum of finite excitation subspaces with smooth wave functions, with decay in infinity as defined for $\Fc_D^S$ and $\Fc_R^S$, will be denoted $\Fc^S$. This space is invariant under the action of the Dirac in-field smeared with a Schwartz function, as well as the electromagnetic in-field \eqref{elm_repr} with a test function as defined above.

For our purposes we shall need further restrictions on the covariance:
\begin{gather}
 B^\frac{1}{2}\Fc^S=\Fc^S\,,\label{elm_BB}\\
 \|B_{\p^2}^{-\frac{1}{2}}[(t\cdot l)^2\p^2]^{-M}\|_{\p^2}<\infty\quad \text{for some}\ M\in\mN\,,\label{elm_BBB}
\end{gather}
with $t$ as defined in footnote \ref{int_conv}.
In the first relation it is meant that the operator acts on one of the \mbox{$l$-arguments} related to one of the spaces $\Hc_\ir$ of a function in $\Fc^S$. Examples of operators satisfying all conditions \eqref{elm_B}, \eqref{elm_BB} and \eqref{elm_BBB} are given by
\begin{equation}
 B_{\p^2}=\kappa\,[(t\cdot l)^2\p^2]^{-2n}\,,\quad \kappa>0\,,\quad n=1,2,\ldots\,.
\end{equation}
For the discussion of the operator $(t\cdot l)^2\p^2$ we refer the reader to Appendix in Ref.\,\cite{her98}.

\section{Almost radial gauge}\label{ar}

For the construction of the quantum electromagnetic interaction one needs a quantum potential; that is, we need to be able to smear the potential
\eqref{int_pot} with a vector function $K(x)$ with similar asymptotic behavior as $J(x)$ in the last section, but which need not be conserved. For the free incoming potential $A^\inc$ we use the almost radial gauge constructed in \cite{her22}, and for the retarded potential of the quantum current $j(x)$ we give a tentative definition in Section \ref{firstelm} below. The gauges of both parts are functionals of an auxiliary real Schwartz function $\rho$ on Minkowski space, such that
\begin{equation}\label{ar_rho}
 \int\rho(x)dx=1\,,\quad \int\rho(x)x^\al\,dx=0\,,\quad 1\leq|\al|\leq n\,,
\end{equation}
where $\alpha$ is a multi-index and $n$ is a (large) natural number,\footnote{In fact, one can assume for simplicity that $n=\infty$.} which is equivalent to
\begin{equation}
 \hat{\rho}(0)=\frac{1}{2\pi}\,,\quad (\p^\alpha\hat{\rho}\,)(0)=0\,,\quad 1\leq|\alpha|\leq n\,.
\end{equation}

In this section we briefly summarize the construction of the free almost radial gauge. We define
\begin{equation}
 \zeta(s,l)=\int\frac{\rho(a)}{s-a\cdot l}\,da\,,\qquad
 \eta(s,l)=\int\frac{a\rho(a)}{s-a\cdot l}\,da\,,
\end{equation}
and also denote for later use
\begin{equation}\label{ar_kappa}
 \kappa(x,l)=x\zeta(x\cdot l,l)-\eta(x\cdot l,l)
 =\int \frac{(x-a)\rho(a)}{(x-a)\cdot l}\,da\,.
\end{equation}
Let $K(x)$ be as described above and denote
\begin{align}
 &W(s,l)=\int\delta(s-x\cdot l)K(x)\,dx\,,\label{ar_WK}\\
 &\begin{aligned}
 \dV_K(s,l)&=\dW(s,l)+\eta(s,l)l\cdot\dW(s,l)+\zeta(s,l)\p\big(l\cdot W(s,l)\big)\\
 &=\dW(s,l)+\p_s\big[\eta(s,l)l\cdot W(s,l)\big]
 +\p\big[\zeta(s,l)l\cdot W(s,l)\big]\label{ar_VK}\,,
\end{aligned}
\end{align}
where for the second equality in \eqref{ar_VK} we used properties of functions $\zeta$ and $\eta$, see Appendix C in \cite{her22}.
The profile $V_K(s,l)$ satisfies the demands of the extended algebra and the almost radial gauge is defined by
\begin{equation}\label{ar_def}
 A^\inc(K)=\{V_K,V^\mathrm{q}\}\,,
\end{equation}
see formulas (72) and (57) in \cite{her22}.  In order to apply formula \eqref{elm_repr} we need the Fourier transform (we use the first form in  \eqref{ar_VK})
\begin{equation}\label{ar_V}
\begin{aligned}
 &\ti{\dV}_K(\w,l)=\ti{\dW}(\w,l)\\
 &+\pi\int_\mR\sgn(u)\bigg\{(\p\hat{\rho})(ul)\,l\cdot \ti{\dW}(\w-u,l)
 +i\hat{\rho}(ul)\,\p\Big[l\cdot\ti{W}(\w-u,l)\Big]\bigg\}\,du\,,
\end{aligned}
\end{equation}
where
\begin{equation}\label{ar_FW}
 \ti{W}(\w,l)=\hat{K}(\w l)\,,
\end{equation}
and we used relations (see Appendix C in \cite{her22})
\begin{equation}
 \ti{\zeta}(\w,l)
 =i\pi\sgn(\w)\hat{\rho}(\w l)\,,\quad
 \ti{\eta}(\w,l)
 =\pi\sgn(\w)(\p\hat{\rho})(\w l)\,.
\end{equation}

The commutation relation of the potential in the almost radial gauge follows from \eqref{elm_algV} and \eqref{ar_def}:
\begin{equation}
 \big[A^\inc(K_1),A^\inc(K_2)\big]=i\{V_{K_1},V_{K_2}\}\,.
\end{equation}
We obtain in Appendix \ref{sform} a more explicit form of the commutator function in terms of smearing fields:
\begin{equation}
 \{V_{K_1},V_{K_2}\}=4\pi\int K_1^a(x)\Dc_{ab}(x,y)K_2^b(y)\,dx\,dy\,,
\end{equation}
where
\begin{equation}\label{ar_Dcom}
 \Dc_{ab}(x,y)=g_{ab}D(x-y)+\p^x_a\Fc_b(x,y)-\p^y_b\Fc_a(y,x)
 +\p_a^x\p_b^y\Gc(x,y)\,,
\end{equation}
\begin{align}
 D(x)&=-\frac{1}{8\pi^2}\int\delta'(x\cdot l)\,d^2l
 =\frac{1}{2\pi}\sgn(x^0)\delta(x^2)\,,\label{ar_PJ}\\
 \Fc_b(x,y)&=\frac{1}{8\pi^2}
 \int\delta\big((x-y)\cdot l\big)\kappa_b(x,l)\,d^2l\,,\\
 \Gc(x,y)&=\frac{1}{16\pi^2}
 \int\sgn\big((x-y)\cdot l\big)\kappa(x,l)\cdot\kappa(y,l)\,d^2l\,.
\end{align}

\section{Interacting fields}\label{inter}

\subsection{First order Dirac field}

For a Schwartz bi-spinor function $\chi(x)$ the smeared Dirac field is denoted
\begin{equation}\label{idir_def}
 \langle \chi,\psi\rangle=\int\bar{\chi}(x)\psi(x)\,dx
 =(2\pi)^{-2}\int\ov{\hat{\chi}}(p)\hat{\psi}(p)\,dp
 =(2\pi)^{-2}\langle\hat{\chi},\hat{\psi}\rangle\,.
\end{equation}

The first order Dirac field has the form
\begin{equation}\label{idir_def1}
\begin{aligned}
 \langle\chi,\psi^{(1)}\rangle
 &=-e\int \bar{\chi}(x)S^\ret(x-y)dx\,\ga^b\psi^\inc(y)A^\inc_b(y)\,dy\\
 &=\sum_\vep\int
 K_\chi^b(y,\vep q)\,c_\vep(q)\,d\mu_m(q)\,A^\inc_b(y)\,dy\,,
\end{aligned}
\end{equation}
with
\begin{equation}
 K_\chi^b(y, r)
 =-\frac{e}{(2\pi)^\frac{3}{2}}\int\bar{\chi}(x) S^\ret(x-y)\,dx\,\ga^b e^{-ir\cdot y}P(r)\,,\quad r=\vep q\,,\quad q\in H_m\,,
\end{equation}
where $P(r)=P(\vep q)$ is as defined in \eqref{dir_P}. Straightforward calculation gives
\begin{equation}
 \hat{K}_\chi(k, r)=-\frac{e}{(2\pi)^\frac{3}{2}}\,
 \frac{\ov{\hat{\chi}}(r-k)
 (r-\frac{1}{2}\fsl{k}\ga)}{r\cdot k-\frac{1}{2}k^2-i\sgn(r^0-k^0)\,0}P(r)\,,
\end{equation}
and its well-defined restriction to $k=\w l$ is now set into formula \eqref{ar_FW}. In this way one obtains
\begin{equation}
 \ti{W}_\chi(\w,l,r)
 =-\frac{e}{(2\pi)^\frac{3}{2}}\,\ov{\hat{\chi}}(r-\w l)\,
 \frac{r-\frac{1}{2}\w\fsl{l}\ga}{(\w-i0)\,r\cdot l}P(r)\,.
\end{equation}
Then
\begin{align}
 \ti{\dW}_\chi(\w,l,r)
 &=\frac{ie}{(2\pi)^\frac{3}{2}}\,\ov{\hat{\chi}}(r-\w l)\,
 \frac{r-\frac{1}{2}\w\fsl{l}\ga}{r\cdot l}P(r)\,,\\
 \frac{\p}{\p l}\big(l\cdot \ti{W}_\chi(\w,l,r)\big)
 &=\frac{e}{(2\pi)^\frac{3}{2}}\,\ov{\partial\hat{\chi}}(r-\w l)P(r)\,,\\
 l\cdot \ti{\dW}_\chi(\w,l,r)
 &=\frac{ie}{(2\pi)^\frac{3}{2}}\,\ov{\hat{\chi}}(r-\w l)P(r)\,.
\end{align}
Putting these expressions into the formula \eqref{ar_V} one obtains
\begin{equation}\label{idir_V}
\begin{aligned}
 \ti{\dV}_{\!\chi}(\w,l,r)
 &=\frac{ie}{(2\pi)^\frac{3}{2}}\bigg\{\ov{\hat{\chi}}(r-\w l)
 \frac{r-\frac{1}{2}\w\fsl{l}\ga}{r\cdot l}\\
 &+\pi\int_\mR\sgn(u)\p\big[\hat{\rho}(ul)\,\ov{\hat{\chi}}(r-\w l+ul)\big]
 du\bigg\}P(r)\,;
\end{aligned}
\end{equation}
here and in what follows we use a shorthand notation
\begin{equation}
 \p\big[A(p)B(k)\big]=(\p A)(p)B(k)+A(p)(\p B)(k)
\end{equation}
For $\w=0$ one finds
\begin{equation}\label{idir_V0}
\begin{aligned}
 \ti{\dV}_{\!\chi}(0,l,r)
 &=\frac{ie}{(2\pi)^\frac{3}{2}}\bigg\{\ov{\hat{\chi}}(r)
 \frac{r}{r\cdot l}
 +\pi\int_\mR\sgn(u)\p\big[\hat{\rho}(ul)\ov{\hat{\chi}}(r+ul)\big]
 \,du\bigg\}P(r)\\
 =\p\Phi_\chi(l,r)
 &\equiv\frac{\p}{\p l}\bigg\{\frac{-ie}{2(2\pi)^\frac{1}{2}}
 \int\sgn(u)\log|u r\cdot l|\,
 \p_u\big[\hat{\rho}(ul)\ov{\hat{\chi}}(r+ul)\big]\,du\,P(r)\bigg\}\,.
 \end{aligned}
\end{equation}

\begin{thm}\label{idir_psi1}
Let the covariance operator $B$ satisfy conditions \eqref{elm_B} and \eqref{elm_BB}. Then the first order Dirac field $\langle\chi,\psi^{(1)}\rangle$ is an operator mapping $\Fc^S$ to $\Fc^S$.
\end{thm}
\begin{proof}
The function \eqref{idir_V} is smooth in all arguments. Next, we note that if a function $f(x,y)$, $x,y\in \mR^n$, is of fast decrease in variables $x$ and $y$, then it is also of fast decrease in $x$ and $x+y$.
With this observation, a close inspection of the formulas \eqref{idir_V} and \eqref{idir_V0} shows that for each $k,m,n\in\{0,1,\ldots\}$ there exists a fixed finite constant $\sigma\geq1$ such that for each $N\geq1$ we have\footnote{Powers of operators $L$ and $M$ are symbolic, each factor has its own indices. Also, here, and in similar contexts, the arbitrarily large negative exponents $-N$ are treated symbolically: at logically related estimates they do not need to denote the same number.}
\begin{align}
 |\p_\w^kL^mM^n\ti{\dV_\chi}(\w,l,r)|
 &\leq \con\frac{(1+|\w|)^\sigma|r|^\sigma}{(1+|r-\w l|)^N}\label{idir_bV}\\
 |L^mM^n \Phi_\chi(l,r)|
 &\leq\frac{\con}{|r|^N}\label{idir_0}
\end{align}
(note that $|r|\geq m$); the precise value of $\sigma$ does not matter for us, and the multiplicative constants on the rhs depend on all parameters. The first bound above also implies
\begin{align}
  &\big|\p_\w^kL^mM^n\, \ti{V}_{\chi,h}(\w,l,r)\big|
  \leq \con\frac{(1+|\w|)^\sigma|r|^\sigma}{(1+|r-\w l|)^N}\,, \label{idir_w0}\\
  &|L^mM^n R_h(l,r)|=
  \Big|L^mM^n\int\ov{\ti{h}}(\w,l)\ti{\dV_\chi}(\w,l,r)\frac{d\w}{\w}\Big|
  \leq\frac{\con}{|r|^N}\,,\label{idir_h}
\end{align}
with $\ti{V}_{\chi,h}(\w,l)$ as defined in \eqref{elm_Vh}. Estimate \eqref{idir_w0} follows easily from \eqref{idir_bV}. For \eqref{idir_h} we use the second equality in \eqref{elm_RV} and the observation made at the beginning of the proof. The quantities on the lhs of \eqref{idir_0} and \eqref{idir_h} are thus sufficiently bounded, and we only need to comment on \eqref{idir_w0}. For $\vep\w<0$ (the cases `creation-creation' and `annihilation-annihilation') the rhs of \eqref{idir_w0} is bounded by $\con[(1+|\w|)|r|]^{-N}$ for each $N\geq0$. For $\vep\w>0$ (the cases `creation-annihilation' and `annihilation-creation') the bound is integrated with a fast decreasing function of either $r$ or $\w$, which produces on the rhs $\con(1+|\w)^{-N}$ or $\con|r|^{-N}$, respectively. Next, we recall definition \eqref{elm_rV} and note that the operator $P_\ir$ conserves smoothness and fast decay of the function $R_h(l,r)$ (see the explicit form of this operator given in Lemma \ref{cone_proj} in Appendix \ref{cone}). Finally, we also recall definitions \eqref{elm_pV} and \eqref{elm_jV}, and note that $B^{\pm\frac{1}{2}}$ leave $\Fc^S$ invariant (by the assumption \eqref{elm_BB}).
\end{proof}

\subsection{First order electromagnetic field}\label{firstelm}

Here we briefly comment on a possible choice of distribution $\Dc^\ret_{ab}(x,y)$ in equation \eqref{int_pot}, which should replace the naive $g_{ab}D^\ret(x-y)$, so as to ensure extension of a kind of almost radial property to the retarded field. One possibility worth considering would be to split $D^\ret$ into the sum of $\frac{1}{2}D=\frac{1}{2}(D^\ret-D^\adv)$ and $D^s=\frac{1}{2}(D^\ret+D^\adv)$, and replace
\begin{equation}
\begin{aligned}
 g_{ab}\tfrac{1}{2}D(x-y)&\to \tfrac{1}{2}\Dc_{ab}(x,y)\,,\\
 g_{ab}D^s(x-y)&\to g_{ab}D^s(x-y)-
 \p_a^x\Big\{\int \rho(a)\tfrac{1}{2}\log|(x-a)^2|da\, x_bD^s(x-y)\Big\}\,,
\end{aligned}
\end{equation}
where $\Dc_{ab}(x,y)$ is the commutator function \eqref{ar_Dcom}, and $\rho(x)$ is the function introduced in Section \ref{ar}. However, as we are not going to consider higher order terms, we content ourselves with the first order electromagnetic \emph{field}. We conjecture, that for $J$ conserved and of Schwartz behavior one has
\begin{equation}\label{firstelm1}
 A^{(1)}(J)=4\pi\int J(x)D^\ret(x-y)j^\inc(y)dy\,dx\,.
\end{equation}
Moreover, as the current $j^\inc(y)$ does not radiate, we can replace $D^\ret$ by $D^s$ in this formula.
\begin{thm}
 For a Schwartz function $\chi$ and a Schwartz conserved current $J$ the operators $j^\inc(\chi)$ and $A^{(1)}(J)$ map $\Fc^S$ to $\Fc^S$.
\end{thm}
The proof of these statements, which we do not give here in details, may be easily inferred as a byproduct of the proof of Theorem \ref{space_limits} in Section \ref{space} below. For the (infinite) differentiability of the wave function \eqref{space_psipS} in $\pb$ it is sufficient to note that
\[
 \frac{\p}{\p\pb}\frac{1}{(p-q)^2}
 =-\frac{q^0}{p^0}\frac{\p}{\p\qb}\frac{1}{(p-q)^2}\,,
\]
and integrate $\p/\p\qb$ by parts.

\section{Translations}\label{transl}

The algebra of free fields $\psi^\inc(\chi)$ and $A^\inc(J)$, where $\p\cdot J=0$, is invariant under the usual action of the Poincar\'e group. As mentioned above, the translation group is unitarily implementable, that is
\begin{equation}
\begin{aligned}
 \alpha_z(\psi^\inc(\chi))&\coloneqq \psi^\inc(T_z\chi)=U(z)\psi^\inc(\chi)U(-z)\,,\\
 \alpha_z(A^\inc(J))&\coloneqq
 A^\inc(T_zJ)=U(z)A^\inc(J)U(-z)\,,
 \end{aligned}
\end{equation}
where $(T_zf)(x)=f(x-z)$. Also, composite fields, if they are translation  invariantly constructed with the use of these fields (which means, in particular, that $A^\inc$ is always smeared with a conserved current), will satisfy these relations. However, the incoming almost radial gauge $A^\inc(K)$, for non-conserved $K$, depends on an auxiliary function $\rho$. If this dependence is made explicit by writing $A^\inc(K;\rho)$, then the above action gives $\alpha_z(A(K;\rho))=A(T_zK; T_z\rho)$, which has been discussed in \cite{her22}. For the shifted function $T_z\rho$ the condition \eqref{ar_rho} is satisfied with respect to the origin shifted to $z$.
However, this is a change of gauge, which should not have impact on observables. Therefore, we can combine the transformation $\alpha_z$ with the restoration of the initial function $\rho$ and define $\hat{\alpha}_z(A^\inc(K;\rho))=A^\inc(T_zK;\rho)$; for observables, which do not depend on $\rho$, one has $\hat{\alpha}_z=\alpha_z$.

As the total quantum current $j(x)$ should turn out to be such observable, also the total potential \eqref{int_pot} should transform as $\alpha_z(A(K;\rho))=A(T_zK;T_z\rho)$, and then keeping $\rho$ constant as explained above we can translate potential by $\hat{\alpha}_z(A(K;\rho))=A(T_zK;\rho)$.

We expect that all fields of the theory will be transformed covariantly by $\alpha_z$, if $\rho$ is added to the family of test functions. This is easily confirmed for the first order Dirac field \eqref{idir_def1}. Then for all quantities nontrivially dependent on $\rho$ we replace $\alpha_z$ by $\hat{\alpha}_z$. In particular, $\hat{\alpha}_z(\langle\chi,\psi^{(1)}\rangle)=\langle T_z\chi,\psi^{(1)}\rangle$.

\section{IN-asymptotics of the Dirac field}\label{asd}

For the investigation of asymptotics of the fields we use translation $\hat{\alpha}_z$, as discussed in the previous section.

For the derivation of `in'-asymptotics of the massive field $\psi(x)$ we proceed as follows. In the first step the field is smeared with a Schwartz spinor test function $\bar{\chi}(x)$, with the support of the Fourier transform contained in $p^2\geq\vep^2$ for some $\vep>0$ (both future and past parts are allowed). The resulting field is translated with the use of $\hat{\alpha}_z$ by $z=-\la v$, $\la>0$, $v\in H_1$, and then integrated over $v$ with $d\mu(v)=(v^0)^{-1}d^3\mathbf{v}$.  The limit $\la\to\infty$ gives the past asymptotic behavior of the field. It is easy to see that the smearing thus described amounts to
\begin{equation}\label{asd_sm}
 \int \hat{\alpha}_{-\la v}(\langle\chi,\psi\rangle)\,d\mu(v)
 =\int \bar{\chi}(x+\la v)\psi(x)\,dx \,d\mu(v)
 =\int F(\la p)\,\ov{\hat{\chi}}(p)\hat{\psi}(p) \,dp\,,
\end{equation}
where for $w^2>0$
\begin{equation}\label{asd_D}
\begin{aligned}
 F(w)&=(2\pi)^{-2}\int e^{i w\cdot v}d\mu(v)
 =i4\pi D^{(-)}_1(w)\\
 &=\frac{1}{2\sqrt{w^2}}\big(Y_1(\sqrt{w^2})-i\sgn(w^0) J_1(\sqrt{w^2})\big)\\
 &=(2\pi)^{-\frac{1}{2}}(\sqrt{w^2})^{-\frac{3}{2}}
 e^{i\sgn(w^0)(\sqrt{w^2}+\frac{3}{4}\pi)}\Big(1+O\Big(\frac{1}{\sqrt{w^2}}\Big)\Big)\,;
\end{aligned}
\end{equation}
here we have used the known explicit form of the function $D_1^{(-)}$, in the notation of \cite{sch95} (formula (2.3.37), with a spacetime $x$ changed to  $w$, and the `mass' parameter equal to 1), and properties of Bessel functions. We now want to compensate the geometric decay factor, and also dump asymptotic oscillations, in $F(\la p)$. For this purpose we choose a smooth function $d(\la)$ of compact support in $(0,+\infty)$, such that $\int d(\la) d\la=1$, and define for $\La>0$ and $\bar{m}\in\mR$:
\begin{equation}\label{asd_def}
 \langle \chi,\psi\rangle[\bar{m},d,\La]
 =\int G[\bar{m},d,\La](p)\ov{\hat{\chi}}(p)\hat{\psi}(p)\,dp\,,
\end{equation}
where
\begin{equation}
\begin{aligned}
 G[&\bar{m},d,\La](p)
 =\int\sqrt{2\pi}(\la |\bar{m}|)^\frac{3}{2} e^{-i\sgn(p^0)(\la \bar{m}+\frac{3}{4}\pi)}
 F(\la p)d\Big(\frac{\la}{\La}\Big)\frac{d\la}{\La}\\
 &=\int\Big(\frac{|\bar{m}|}{\sqrt{p^2}}\Big)^\frac{3}{2}
 e^{i s\La \sgn(p^0)(\sqrt{p^2}-\bar{m})}
 \Big(1+O\Big(\frac{1}{s\La\sqrt{p^2}}\Big)\Big)d(s)\,ds\\
 &=2\pi\Big(\frac{|\bar{m}|}{\sqrt{p^2}}\Big)^\frac{3}{2}
 \bigg\{\tilde{d}\Big(\La \sgn(p^0)(\sqrt{p^2}-\bar{m})\Big)
 +O\Big(\frac{1}{\La\sqrt{p^2}}\Big)\bigg\}\,.
\end{aligned}
\end{equation}
For each choice of parameters in square brackets, and $p$ restricted to the support of $\ov{\hat{\chi}}(p)$, the function $G[\bar{m},d,\La](p)$ is smooth and bounded, and all its partial derivatives of order $n$ are bounded by $\con|p|^n$ (for fixed parameters in square brackets). Therefore,  $G[\bar{m},d,\La](p)\ov{\hat{\chi}}(p)$ is a Schwartz function, and the operators \eqref{asd_def}, for $\psi=\psi^\inc$ or $\psi=\psi^{(1)}$,  map $\Fc^S$ to $\Fc^S$ (by Theorem \ref{idir_psi1} for $\psi^{(1)}$).

\begin{thm}
For each $\Phi\in\Fc^S$ one has
\begin{align}
 \slim_{\La\to\infty}\,\langle\chi,\psi^\inc\rangle[m,d,\La]\Phi
 &=\langle\chi,\psi^\inc\rangle\Phi\,,\label{asd_limin}\\
 \slim_{\La\to\infty}\,\langle\chi,\psi^\inc\rangle[\bar{m},d,\La]\Phi&=0\,,
 \quad \bar{m}\neq m\,.\label{asd_limfin}
\end{align}
If the covariance operator $B$ satisfies \eqref{elm_B}, \eqref{elm_BB}, and \eqref{elm_BBB}, then
\begin{equation}
 \slim_{\La\to\infty}\,\langle\chi,\psi^{(1)}\rangle[\bar{m},d,\La]\Phi=0\,,\quad \bar{m}\in\mR\,.\label{asd_lim1}
\end{equation}
\end{thm}
\begin{proof}
For \eqref{asd_limin} and \eqref{asd_limfin} we note that in these cases one has $\sqrt{p^2}=m$ in the function $G[\bar{m},d,\La](p)$, and the thesis follows immediately.

For the sake of the proof of \eqref{asd_lim1}, we simplify notation of $G$ by suppressing arguments $\bar{m}$ and $d$, and we write $G[\bar{m},d,\La](p)=G_\La(p)$. We first note that the function $\ti{\dV_\La}(\w,l,r)$ corresponding to $\langle\chi,\psi^{(1)}\rangle[\bar{m},d,\La]$ is obtained from \eqref{idir_V} by the substitution
$\ov{\hat{\chi}}(p)\to \ov{\hat{\chi}}(p)G_\La(p)$. In particular, under the integral in \eqref{idir_V} we have to replace
\begin{equation}\label{asd_repl}
\begin{aligned}
 \p\big[\hat{\rho}(ul)\ov{\hat{\chi}}(r-\w l+ul)\big]\to \
 &\p\big[\hat{\rho}(ul)\ov{\hat{\chi}}(r-\w l+ul)\big]G_\La(r-\w l+ul)\\
 &+\hat{\rho}(ul)\ov{\hat{\chi}}(r-\w l+ul)\,\p G_\La(r-\w l+ul)\,.
\end{aligned}
\end{equation}
Using the relation $\dsp\frac{dG_\La(r-\w l+ul)}{du}=l\cdot\p G_\La(r-\w l+ul)$ and the fact that inside each part of the light cone $G_\La(p)$ depends only on $p^2$, one finds that
\begin{equation}\label{asd_dG}
 \p G_\La(r-\w l+ul)=\frac{r-\w l+ul}{r\cdot l}\,\frac{d}{du}G_\La(r-\w l+ul)\,.
\end{equation}
We use this representation in the second term of the rhs of \eqref{asd_repl}, and integrate this term in the integral in the modified \eqref{idir_V} by parts with respect to $u$. In this way one finds that $\ti{\dV_\La}(\w,l,r)=\ti{\dV^1_\La}(\w,l,r)+\ti{\dV^2_\La}(\w,l,r)$, where
\begin{align}
&\begin{aligned}\label{asd_V1}
 &\ti{\dV^1_\La}(\w,l,r)=\frac{ie}{2\sqrt{2\pi}}
 \int G_\La(r-\w l+ul)\sgn(u)
 \bigg\{\p\Big[\hat{\rho}(ul)\ov{\hat{\chi}}(r-\w l+ul)\Big]\\
 &\hspace{9em}-\p_u\Big[\frac{r-\w l+ul}{r\cdot l}
 \hat{\rho}(ul)\ov{\hat{\chi}}(r-\w l+ul)\Big]\bigg\}\,du\, P(r)\,,
\end{aligned}\\
 &\ti{\dV^2_\La}(\w,l,r)=\frac{ie}{(2\pi)^\frac{3}{2}}\w\,
 G_\La(r-\w l)\ov{\hat{\chi}}(r-\w l)
 \frac{l-\frac{1}{2}\fsl{l}\ga}{r\cdot l}P(r)\,.\label{asd_V2}
\end{align}
The rest of the proof is based on the fact that on the support of $\ov{\hat{\chi}}(p)$ the function $G_\La(p)$ is bounded, and for $\La\to\infty$ tends pointwise to zero almost everywhere in $p$. For $\ti{\dV^2_\La}(\w,l,r)$ we have
\begin{equation}\label{asd_2h}
 \big|\ti{V^2_{\La,h}}(\w,l,r)\big|=\big|\w^{-1}\ti{\dV^2_\La}(\w,l,r)\big|
 \leq |G_\La(r-\w l)|\frac{\con\,|r|^2}{(1+|r-\w l|)^N}\,.
\end{equation}
Arguments similar to those used for \eqref{idir_w0}, together with vanishing of $G_\La$, give now the thesis for this part. Next, using the observation made at the beginning of the proof of Theorem \ref{idir_psi1}, we estimate
\begin{equation}\label{asd_1h}
 \big|\ti{\dV^1_\La}(\w,l,r)\big|
 \leq\con\int\frac{|G_\La(r-\w l+ul)|\,du}{(1+|u|)^N}\,
 \frac{|r|^2}{(1+|r-\w l|)^N}\,,
\end{equation}
which guarantees the thesis for $\ti{V^1_{\La,h}}(\w,l,r)$ in the region $|\w|\geq1$. In the region $|\w|\leq1$ one needs
\begin{equation}\label{asd_pwV1}
\begin{aligned}
 &\p_\w \ti{\dV^1_\La}(\w,l,r)=\frac{ie}{(2\pi)^\frac{3}{2}} G_\La(r-\w l)
 \bigg\{(\p\ov{\hat{\chi}})(r-\w l)
 +\p_\w\Big[\frac{r-\w l}{r\cdot l}
 \ov{\hat{\chi}}(r-\w l)\Big]\bigg\}P(r)\\
 &+\frac{ie}{2\sqrt{2\pi}}
 \int G_\La(r-\w l+ul)\sgn(u)
 \bigg\{\p\Big[(\p_u\hat{\rho})(ul)\ov{\hat{\chi}}(r-\w l+ul)\Big]\\
 &\hspace{9em}-\p_u\Big[\frac{r-\w l+ul}{r\cdot l}
 (\p_u\hat{\rho})(ul)\ov{\hat{\chi}}(r-\w l+ul)\Big]\bigg\}\,du\, P(r)\,,
\end{aligned}
\end{equation}
where $\p_\w$, in action on functions depending on $r-\w l+tl$ in \eqref{asd_V1}, has been changed to $-\p_t$ and integrated by parts. It is now clear that $|\p_\w \ti{\dV^1_\La}(\w,l,r)|$ is estimated by a sum of terms of the types appearing on the rhs of \eqref{asd_2h} and \eqref{asd_1h}. This completes the estimation of $\ti{V}_{\La,h}(\w,l)$ needed for the thesis.

Next, we consider
\begin{equation}
\begin{aligned}
 \ti{\dV_\La}(0,l,r)&=\ti{\dV^1_\La}(0,l,r)=\frac{ie}{2\sqrt{2\pi}}
 \int G_\La(r+ul)\sgn(u)
 \bigg\{\p\Big[\hat{\rho}(ul)\ov{\hat{\chi}}(r+ul)\Big]\\
 &-\p_u\Big[\frac{r+ul}{(r+ul)\cdot l}
 \hat{\rho}(ul)\ov{\hat{\chi}}(r+ul)\Big]\bigg\}\,du\, P(r)=\p\Phi_\La(l,r)\,,
\end{aligned}
\end{equation}
where the denominator in the second line is written as $(r+ul)\cdot l$ in order to extend $\ti{\dV_\La}(0,l,r)$ to a neighborhood of the cone with the preservation of its orthogonality to $l$ and homogeneity of degree $-1$. The existence of $\Phi_\La(l)$ is guaranteed by \eqref{idir_V0}. As
$B_{\p^2}^{-\frac{1}{2}}$ satisfies assumption \eqref{elm_BBB}, it is sufficient to consider
\begin{equation}
 \|[(t\cdot l)^2\p^2]^M\Phi_\La(.,.)\|
 =\|[(t\cdot l)^2\p^2]^{M-1}(t\cdot l)^2\p\cdot\ti{\dV_\La}(0,.\,,.)\|\,.
\end{equation}
The action of $\p$ and of $\p^2$ above yields well-defined results on the cone, as described in Appendix \ref{cone}.
Again, all $l$-derivatives acting on $G_\La$ in $\ti{\dV_\La}(0,l,r)$ may be integrated by parts as before, and vanishing of this norm follows on similar lines as above.

Finally, we consider $R_h(\dV_\La)$. First, we note that for \eqref{asd_V2} we have
\begin{align}
 R_h(\dV^2_\La)(l,r)
 &=\frac{e}{(2\pi)^\frac{3}{2}}\int
 G_\La(r-\w l)\ov{\hat{\chi}}(r-\w l)\frac{l-\frac{1}{2}\fsl{l}\ga}{r\cdot l}\ov{\ti{h}}(\w,l)d\w\,P(r)\,,\\
 |R_h(\dV^2_\La)(l,r)|
 &\leq\con\int\frac{|G_\La(r-\w l)|\,d\w}{(1+|\w|)^N}\ \frac{1}{|r|^N}\,,
\end{align}
so $\|R_h(\dV^2_\La)\|\to0$ for $\La\to\infty$.
For $R_h(\dV^1_\La)$ one obtains similar conclusion with the use of the second equality in \eqref{elm_RV} and the estimate of $\ti{V}_{\La,h}(\w,l)$ derived above. As the operator $B^\frac{1}{2}P_\ir$ is bounded, the thesis follows.
\end{proof}

\section{Spacelike and past timelike asymptotics of the electromagnetic current and field}\label{space}

We start by noting that the matrix element $(\Phi_1,j^\inc(x)\Phi_2)$, $\Phi_i\in\Fc^S$, is a smooth function obtained with the use of \eqref{dir_curr}. The stationary phase expansion, Corollary  \ref{packet_ascur} in Appendix \ref{packet}, gives its asymptotic behavior, and we find
\begin{equation}\label{space_weakj}
\begin{aligned}
 (\Phi_1&,j^\inc(x)\Phi_2)
 =\theta(x^2-1) \Big(\frac{m}{|\la|}\Big)^3\bigg\{u\,(\Phi_1,\rho_D(mu)\Phi_2)\\
 &+i\sgn(\la)e\sum_{\vep}\vep e^{-i2m\vep \la}
 (\Phi_1,:\bar{c}_{-\vep}(mu)\ga c_\vep(mu):\Phi_2)+O(|x|^{-4})\bigg\}\\
 &+O((1+|x|)^{-N})\,,
\end{aligned}
\end{equation}
where for $x^2\geq1$ we have denoted $x=\la u$, $\la\in\mR$, $u\in H_1$, and where $\rho_D(p)$ is the operator of momentum-density of charge \eqref{dir_dench}. The matrix element vanishes fast in spacelike directions. The `creation-creation' and `annihilation-annihilation' contributions to the leading term in timelike directions oscillate rapidly, and after appropriate averaging also vanish rapidly. The above expansion could also serve for the analysis of the retarded field, but we shall use another method, which will produce limits in strong topology.

We want to investigate the asymptotic limit of the electromagnetic current and field in the spirit of Eqs. \eqref{int_ascur} and \eqref{int_tail}, but with the representation of the field in the form $A(J)$, where $J$ is a conserved test current. Also, the test current $J$ for $A(J)$, and the test field $\chi$ for $j(\chi)$, are not assumed to be of compact support, but rather are Schwartz functions supported outside the future light cone $V_+$.
We have seen that the first order Dirac field $\psi^{(1)}$ decays faster than $\psi^\inc$ in past timelike infinity. We conjecture that this will remain true for higher orders, and that in consequence the first and higher order corrections to the current vanish faster than the current of free Dirac particles. If this is confirmed, then the asymptotics of the electromagnetic field up to the first order should not be modified by higher orders.

We want to know whether the limit in the sense described above will depend on the choice of the central point. Thus we choose a current $J(x)$ and a field $\chi(x)$ with the support outside $V_+$, any vector $z$ in spacetime, and for all $R>0$ denote
\begin{equation}\label{space_def}
\begin{aligned}
 &J_{zR}(x)=R^{-3}J\Big(\frac{x-z}{R}\Big)\,,\quad
 A_{zR}(J)=A(J_{zR})\,,\\
 &\chi_{zR}(x)=R^{-1}\chi\Big(\frac{x-z}{R}\Big)\,,\quad
 j_{zR}(\chi)=j(\chi_{zR})\,.
\end{aligned}
\end{equation}
The Fourier transforms of the scaled test fields are
\begin{equation}\label{space_scale}
 \hat{J}_{zR}(p)=e^{iz\cdot p}R\hat{J}(Rp)\,,\quad
 \hat{\chi}_{zR}(p)=e^{iz\cdot p}R^3\hat{\chi}(Rp)\,.
\end{equation}
We recall the definition \eqref{elm_VJ} of the function $V(s,l)$ determined by $J(x)$, and then we find that the function $V_{zR}(s,l)$ determined by $J_{zR}(x)$ satisfies
\begin{equation}\label{space_scaleV}
 V_{zR}(s,l)=V\Big(\frac{s-z\cdot l}{R}\Big)\,,\quad
 \ti{V}_{zR}(\w,l)=e^{i\w z\cdot l}R\ti{V}(R\w,l)\,.
\end{equation}

\begin{thm}\label{space_limits}
Let $\Psi$ be any vector in $\Fc^S$. With the above notation one has:
\begin{itemize}
\item[(i)] For $A^\inc$, with $V_\ir(l)=-P_\ir V(0,l)$:
\begin{align}
 \wlim_{R\to\infty}\, A^\inc_{zR}(J)\Psi
 &=[a_\ir(iB^{\frac{1}{2}}V_\ir)+a^*_\ir(iB^{\frac{1}{2}}V_\ir)]\Psi\,,\label{space_spacein}\\
 \lim_{R\to\infty}\|A^\inc_{zR}(J)\Psi\|^2 &=\big(\Psi,\|\ti{V}\|_\reg^2
 +[a_\ir(iB^{\frac{1}{2}}V_\ir)+a^*_\ir(iB^{\frac{1}{2}}V_\ir)]^2\,\Psi\big)\,.\label{space_spacein_norm}
\end{align}
If $J$ is supported in $V_-$, then $V_\ir=0$, and if its support is spacelike, then $V_\ir(l)=\p\Phi(l)$, where
\begin{equation}\label{space_Phi}
 \Phi(l)=-\frac{1}{2}\int\sgn(x\cdot l)\frac{x\cdot J(x)}{x^2}\,dx\,.
\end{equation}
\item[(ii)] For $A^{(1)}$:
\begin{equation}\label{space_spaceret}
 \slim_{R\to\infty}A^{(1)}(J_{zR})\Psi
 =\int C(p)\cdot p\, \rho_D(p)\,d\mu_m(p)\,\Psi\,,
\end{equation}
where
\begin{equation}\label{space_Cp}
 C(p)=\int\frac{J(x)}{\sqrt{(p\cdot x)^2-p^2x^2}}\,dx
\end{equation}
\item[(iii)] For $j^\inc$:
\begin{equation}\label{space_spacejin}
 \slim_{R\to\infty}j^\inc(\chi_{zR})\Psi
 =\int\bigg\{\int_\mR \chi(sp)ds\bigg\}p\,\rho_D(p)\,d\mu(p)\,\Psi\,,
\end{equation}
which agrees with the first term of the expansion \eqref{space_weakj}.
\end{itemize}
The algebra of limit operators (i) -- (iii) is commutative.
\end{thm}
\begin{proof}
(i) We write $A^\inc(J_{zR})=A^\inc_\reg(J_{zR})+A^\inc_\ir(J_{zR})$, with the $\reg/\ir$ part defined by $a^\#_\reg$/$a^\#_\ir$ in \eqref{elm_repr}, with \eqref{elm_aaa}, respectively. As $J$ is of fast decrease, so is also $V(s,l)$. Therefore,  $\ti{\dV}(0,l)=0$ and $\ti{V}_h(\w,l)=\ti{V}(\w,l)$, and the same is true for $\ti{V}_{zR}$. A simple change of integration variable shows that \mbox{$\|\ti{V}_{zR}\|_\reg=\|\ti{V}\|_\reg$}, the norm of the product \eqref{elm_scalF}. Similarly, for $\ti{V}_1(\w,l)$ smooth on $[0,\infty)\times C_+$ and of Schwartz behavior in infinity, one finds $(\ti{V}_1,\ti{V}_{zR})\to0$ for $R\to\infty$, and in consequence $\|a_\reg(\ti{V}_{zR})\Psi\|\to0$. Together, these facts imply:
\begin{equation}
 \wlim_{R\to\infty}\, A^\inc_\reg(J_{zR})\Psi=0\,,\quad \text{and}\quad  \lim_{R\to\infty}\|A^\inc_\reg(J_{zR})\Psi\|^2 =\|\ti{V}\|_\reg^2\,.
\end{equation}
The proof of formulas \eqref{space_spacein} and \eqref{space_spacein_norm} will be now completed if we show that
\begin{equation}
 \|B^\frac{1}{2}P_\ir[R_h(\dV_{zR})+V(0,.)]\|_\ir\to 0\quad  (R\to\infty)\,,\label{space_lim}\,.
\end{equation}
As $B^\frac{1}{2}P_\ir$ is a bounded operator, it is sufficient to estimate the norm of the vector in square brackets above. For this we note that
\begin{equation}
 R_h(\dV_{zR})(l)+V(0,l)
 =\int\Big[1-\ov{\ti{h}(R^{-1}\w,l)}e^{iR^{-1}\w l\cdot z}\Big]
 \ti{V}(\w,l)\,d\w\,,
\end{equation}
which is bounded and tends to zero pointwise, which ends the proof of this part.

To prove statements on $V_\ir$ we note that $V(0,l)=\int\delta(x\cdot l)J(x)\,dx$, so $V_\ir=0$ for $J$ with support in $V_-$. Next, for $J$ with spacelike support, we use the second formula in \eqref{cone_pfromf}, which may be written as
\begin{equation}\label{space_pv}
 \Phi(l)
 =\lim_{q\to l}\Phi(q)\,,\quad
 \Phi(q)=\frac{1}{4\pi}\int\frac{q\cdot[- V(0,l')]}{q\cdot l'}\,d^2l'\,,
\end{equation}
where the limit goes over vectors $q$ in $V_+$. One easily calculates
\[
 \int\frac{\delta(x\cdot l')}{q\cdot l'}\,d^2l'
 =\frac{2\pi\theta(-x^2)}{\sqrt{(q\cdot x)^2-q^2x^2}}\,.
\]
Taking into account that the support of $J$ is spacelike, one finds
\begin{equation}
 \Phi(q)=-\frac{1}{2}\int\frac{q\cdot J(x)}{\sqrt{(q\cdot x)^2-q^2x^2}}\,dx
 =-\frac{1}{2}\int\frac{q\cdot x}{\sqrt{(q\cdot x)^2-q^2x^2}}
 \frac{x\cdot J(x)}{x^2}\,dx\,,
\end{equation}
where to obtain the second equality, for $x^2<0$ we denote  $\hat{x}=x/\sqrt{-x^2}$, note the identity
\begin{equation}
 \frac{q}{\sqrt{(q\cdot x)^2-q^2x^2}}
 =\p_x\log\big(q\cdot\hat{x}+\sqrt{(q\cdot\hat{x})^2+q^2}\big)
 +\frac{q\cdot x\, x}{x^2\sqrt{(q\cdot x)^2-q^2x^2}}\,,
\end{equation}
and integrate $\p_x$ by Gauss' theorem. One can now perform the limit $q\to l$, which gives the desired formula.

(ii) We denote by $j_{\vep_1\vep_2}$ this part of the Dirac current $j$, which is formed with $\bar{c}_{\vep_1}$ and $c_{\vep_2}$, and then for $A^{(1)}_{\vep_1\vep_2}(x)=4\pi\int D^s(x-y)j_{\vep_1\vep_2}(y)\,dy$ (see \eqref{firstelm1} and remark following this formula), using \eqref{dir_curr} one finds
\begin{equation}\label{space_vepvep}
\begin{aligned}
 &A^{(1)}_{\vep_1\vep_2}(J_{zR})\\
 &=-\frac{e}{\pi}\int e^{iz\cdot(\vep_1 p-\vep_2 q)}
 \frac{R\hat{J}^a(R(\vep_1p-\vep_2q))}{(\vep_1p-\vep_2q)^2}
 :\bar{c}_{\vep_1}(p)\ga_a c_{\vep_2}(q):\,
 d\mu_m(p)\,d\mu_m(q)\,.
\end{aligned}
\end{equation}
For $\vep_1=-\vep_2=\vep$, one has
\begin{equation}
 \int\Big|\frac{R\hat{J}(\vep R(p+q))}{(p+q)^2}\Big|^2\,d\mu_m(p)\,d\mu_m(q)
 \leq R^2\int\frac{d^3\pp\,d^3\qq}{R^{N+2}(p^0+q^0)^{N+2}}\leq\frac{\con}{R^N}
\end{equation}
for each $N$, so $\|A^s_{\vep,-\vep}(J_{zR})\Psi\|\to 0$ for $R\to\infty$.

For $\vep_1=\vep_2=\vep$ the wave function of the vector $A^{(1)}_{\vep\vep}(J_{zR})\Psi$ is composed of finite collection of $n$-excitation functions of the form
\begin{equation}\label{space_psipS}
 \psi_R(p, S)=-\frac{e}{\pi}\int e^{i\vep z\cdot(p-q)}
 \frac{R\hat{J}^a(\vep R(p-q))}{(p-q)^2}P(p)
 f_a(q, S)\,d\mu_m(q)\,,
\end{equation}
where $P(p)$ is as defined in \eqref{dir_P}, $f_a(q,S)=\ga_a[c_+(q)\Psi_n](S)$, $S$ stands for all remaining arguments, and $f$ is of Schwartz behavior. This implies that for each $N$ one has $|f(q,S)|\leq\con_N F(S)/(1+|\qb|)^N$, where $F(S)$ is of fast decrease. We change integration variable $\qb$ to $\rr=R(\pb-\qb)$, and then
 $\psi_R(p, S)=\int \phi_R(p,S,\rr)\,d^3\rr$, where
\begin{equation}
 \phi(p,S,\rr)=-\frac{em}{\pi}e^{iz\cdot(p-q)}
 \frac{\hat{J}^a(R(p^0-q^0, \rr))}{R^2(p-q)^2q^0}
 P(p)f_a(q, S)\,,
\end{equation}
where $q$ is expressed in terms of $p$ and $\rr$. Simple algebra leads to the bound $R^2|(p-q)^2|\geq 4m^2|\rr|^2/(p^0+q^0)^2$, which results in
\begin{equation}
 |\phi_R(p,S,\rr)|
 \leq\frac{\con (p^0)^2F(S)}{|\rr|^2(1+|\rr|)^N(1+||\pb|-R^{-1}|\rr||)^N}
 \leq\frac{\con (p^0)^2F(S)}{|\rr|^2(1+|\rr|)^N}\,.
\end{equation}
The rhs bound implies that for the point-like limit of $\psi_R(p,S)$ ($R\to\infty$) one can go under the integral sign. On the other hand, the first bound above shows that
\begin{equation}
 |\psi_R(p,S)|\leq\int_0^\infty\frac{\con\,(p^0)^2F(S)}{(1+r)^N(1+||\pb|-R^{-1}r|)^N}\,dr
 \leq \frac{\con\, F(S)}{(1+|\pb|)^N}
\end{equation}
---to show this consider the integration regions $t\gtrless R|\pb|/2$ separately. Therefore, to find the strong Hilbert space limit of $\psi_R(p,S)$ it is sufficient to find point-like limit of $\phi_R(p,S,\rr)$. This is easily calculated, and one obtains
\begin{equation}\label{space_limps}
 \lim_{R\to\infty}\psi_R(p,S)
 =\frac{em}{\pi}\int\Big[\rr^2-\Big(\frac{\pb\cdot\rr}{p^0}\Big)^2\Big]^{-1}
 \hat{J}^a\Big(\frac{\pb\cdot\rr}{p^0},\rr\Big)\,\frac{d^3\rr}{p^0}\,P(p)f_a(p,S)\,.
\end{equation}
It is easy to see that the integral in the above expression may be written as a four-dimensional integral
\begin{multline}
 \int \hat{J}^a(r)\delta(r\cdot p)\frac{dr}{-r^2}
 =\frac{1}{2\pi}
 \int_\mR\bigg\{\int\hat{J}^a(r)e^{isr\cdot p}\frac{dr}{-r^2}\bigg\}\,ds\\
 =(2\pi)^2\int J^a(x)\bigg\{\int_\mR D^s(sp-x)\,ds\bigg\}\,dx
 =\pi\int\frac{J^a(x)\,dx}{\sqrt{(p\cdot x)^2-p^2x^2}}\,,
\end{multline}
where we have used the Fourier representation, and the explicit form of $D^s(x)$. We have shown that
\begin{equation}
 \lim_{R\to\infty}\psi_R(p,S)
 =em\int\frac{J^a(x)\,dx}{\sqrt{(p\cdot x)^2-p^2x^2}}P(p)f_a(p,S)\,.
\end{equation}
Noting that $P(\vep p)\ga P(\vep p)=\vep P(\vep p) p/m$, one obtains the thesis.

(iii) For $j^\inc$ we have
\begin{equation}
\begin{aligned}
 &j^\inc_{\vep_1\vep_2}(\chi_{zR})\\
 =\frac{e}{(2\pi)^3}&\int e^{iz\cdot(\vep_1 p-\vep_2 q)}
 R^3\hat{\chi}(R(\vep_1p-\vep_2q))
 :\bar{c}_{\vep_1}(p)\ga c_{\vep_2}(q):\,
 d\mu_m(p)\,d\mu_m(q)\,,
\end{aligned}
\end{equation}
and the proof is very similar to that for $A^{(1)}(J_{zR})$. Effectively, the difference is that the rhs of \eqref{space_limps} is replaced by
\begin{equation}
 \frac{em}{(2\pi)^2}\int
 \hat{\chi}\Big(\frac{\pb\cdot\rr}{p^0},\rr\Big)\,
 \frac{d^3\rr}{p^0}\,P(p)f_a(p,S)
 =em\int_\mR\chi(sp)\,ds\,P(p)f_a(p,S)\,,
\end{equation}
which ends the proof of (iii). The commutativity of \eqref{space_spacein}, \eqref{space_spaceret} and \eqref{space_spacejin} is a consequence of the commutativity of $a_\ir^\#(iB^{\frac{1}{2}}V_\ir)$ and $\rho_D(p)$ (for the commutativity of the operators $a_\ir(iB^{\frac{1}{2}}V_\ir)+a^*_\ir(iB^{\frac{1}{2}}V_\ir)$ note that their arguments are purely imaginary).

\end{proof}

\section{Past null asymptotics of electromagnetic field}\label{null}

Here we consider past null asymptotics for $A(J)$, where $J$ is conserved and of Schwartz type. We follow the method introduced by Buchholz \cite{buch77}. We shift $A(J)$ into the past null directions by $-r k$, $r>0$, $k\in C_+$, $k^0=1$, multiply by $-2r$, integrate over $r$ with $d(r/R)/R$, with $d$ as in \eqref{asd_def}, average over angles of $\kb$, and finally take the time derivative of the resulting smearing current. This procedure has a clear interpretation of the past null limit, and produces $A(J^d_R)$ with
\begin{equation}
 J^d_R(x)=-2\int r\p_0J(x+r k)d\Big(\frac{r}{R}\Big)\frac{dr}{R}\,\frac{d\W_\kb}{4\pi}\,,
\end{equation}
which gives the Fourier transform
\begin{equation}
 \hat{J}^d_R(p)=\hat{J}(p)\frac{p^0}{|\pb|}F_R(p)\,,\quad
 F_R(p)=2\pi\big[\ti{d}(-R(p^0-|\pb|))-\ti{d}(-R(p^0+|\pb|))\big]\,.
\end{equation}
If we denote by $V^d_R(s,l)$ the $V$ function obtained from $J^d_R$, then
\begin{equation}\label{null_Vla}
 \ti{V}^d_R(\w,l)=\hat{J}^d_R(\w l)
 = \big(1-2\pi \ti{d}(-2R\w l^0)\big)\ti{V}(\w,l)\,.
\end{equation}

\begin{thm}
 Let $J(x)$ and $J^d_R(x)$ be as defined above. Then for each $\Psi\in\Fc^S$ one has
\begin{align}
 &\slim_{R\to\infty} A^\inc(J^d_R)\Psi=A^\inc(J)\Psi\,,\\
 &\slim_{R\to\infty} A^{(1)}(J^d_R)\Psi=0\,.
\end{align}
\end{thm}
\begin{proof}
With $\ti{V}^d_R$ given by \eqref{null_Vla}, the case of $A^\inc$ is trivial. In the case of $A^{(1)}$ we have
\begin{equation}
\begin{aligned}
 &A^{(1)}_{\vep_1\vep_2}(J^d_R)
 =-\frac{e}{\pi}\int
 \frac{\hat{J^d_R}(\vep_1p-\vep_2q)}{(\vep_1p-\vep_2q)^2}
 :\bar{c}_{\vep_1}(p)\ga c_{\vep_2}(q):\,
 d\mu_m(p)\,d\mu_m(q)\,.
 \end{aligned}
\end{equation}
The proof is now a rather obvious modification of the proof of Theorem \ref{space_limits} (ii). Consider $\hat{J^d_R}(\vep_1p-\vep_2q)$. For $\vep_1=-\vep_2$, one has to note that $|\pb+\qb|^{-2}$ is square integrable in the origin, and for $\vep_1=\vep_2$, that $|p^0-q^0|/|\pb-\qb|\leq1$. The pointwise vanishing of $F_R(\vep_1p-\vep_2q)$ almost everywhere gives now the thesis.

\end{proof}

\section{Conclusions and discussion}\label{diss}

The purpose of the article was to give preliminary steps for a construction of QED in a physical gauge. This is assumed to mean that the representation space is a Hilbert space (no indefinite metric), and the interacting fields are properly represented. This has been achieved in the first perturbation order. Treatment of UV problem is a subject for further investigation.

The construction is based on the extension of the usual free electromagnetic field algebra to the infrared-singular fields, with its representation obtained earlier. Another key element is the use of the almost radial gauge of the electromagnetic potential, constructed recently. The formalism thus constructed contains in a nontrivial way long-range, nonlocal degrees of freedom. Translations are implemented in the representation by a unitary group, with energy-momentum spectrum in the forward light cone, but there is no vacuum. However, there do exist certain states on which the representation is built in a Fock way. The energy-momentum content of such states may be arbitrarily low (which is illustrated in Appendix \ref{omega}).

We summarize the results of the investigation.

The first order Dirac field has been constructed as an operator in the representation space (which has not been achieved in standard local formulations). Moreover, the appropriately defined past timelike limit of the Dirac field recovers the free incoming field, while the first correction addition vanishes (decays faster). The use of the almost radial gauge seems decisive for this fact. Thus, there is no `dressing' of the incoming electrons, which is necessary in the local formulations.
We conjecture that this decay will be confirmed in higher orders of perturbation. Consequently, we also expect that the electromagnetic current of the incoming particles is asymptotically equal to that formed by the free incoming Dirac field. If this is confirmed, then the following further results, here based on the first order, will find full justification.

The electromagnetic field has been investigated in asymptotic limits. The spacelike asymptotic behavior, as defined in \eqref{int_tail}, is of the Coulomb rate, but in contrast to the local theory, produces a limit operator (and not only a superselection label). This limit field is a sum of two contributions: the free incoming field limit and the Coulomb field due to incoming particles. The scaling used for this limit, Eqs. \eqref{space_scale} and \eqref{space_scaleV}, clearly shows that this may be also interpreted as a low-energy limit. In this article we work on incoming fields, but an analogous picture may be obtained with the use of  outgoing fields, and the spacelike limit expressed in terms of them must agree with that obtained here. Moreover, we have shown that this limit does not depend on the choice of the central point for scaling. Thus, for each smearing current $J(x)$ of Schwartz type, with support in $x^2<0$, it represents a conserved quantity, given here in precise mathematical terms, which we write down here once more in terms of the incoming quantities:
\begin{equation}
 A^0(J)=
 \big[a_\ir(iB^{\frac{1}{2}}\p\Phi)+a^*_\ir(iB^{\frac{1}{2}}\p\Phi)\big]
 +\int C(p)\cdot p\, \rho_D(p)\,d\mu_m(p)\,,
\end{equation}
where $\Phi(l)$ and $C(p)$ are linear functionals of $J(x)$, as defined in \eqref{space_Phi} and \eqref{space_Cp}. Equality of this operator to the analogous one formed in terms of outgoing quantities may be regarded as a mathematically precise formulation, in our quantum formalism, of the classical `matching property' \cite{her95} (see also \cite{her17}). We note that the operators $A^0(J)$ form an Abelian subalgebra of the algebra of the theory, but are not in the center, which is trivial (the representation is irreducible). In particular, they do not commute with $A^\inc(J')$, smeared with $J'$ with a nontrivial past timelike tail of the form $x\rho(x)$ (admitted in the theory, as explained in Section \ref{elm}). These more general fields have not been considered in our investigation of asymptotic limits, but they certainly appear in the formalism, see the construction of the first order Dirac field $\psi^{(1)}$, where the smearing function \eqref{idir_V} has a nontrivial characteristic \eqref{idir_V0}.\footnote{See also the discussion of the radiation by external current in \cite{her08}. The use of the extended algebra avoids the usual `infrared catastrophe' in this setting, and allows the explicit construction of the scattering operator in terms of an infrared-singular field, see Eq.\ (55) in this reference.}
Another point we would like to stress, is that the limit field $A^0(J)$ is achieved with fluctuations remaining under control, as shown by Theorem~\ref{space_limits}. Thus our representation is not of `infravacuum'-type representations, in which fluctuations are so large in infinity as to mask any addition to the field. At the same time, this limit field is a true quantum variable subject to the usual quantum uncertainties, in contrast to the asymptotic field in the strictly local theory, with `dressed' particles, where it is a c-number (as discussed in Introduction). We leave further more extensive discussion of these questions to future publications.

The `in' limit for the electromagnetic field consists of two regions: the timelike and the null infinity. In the null limit (with Schwartz test currents) one recovers the incoming free field, which is what one should expect. On the other hand, the past timelike limit produces the Coulomb field of the incoming charged particles. We note that this field has been postulated on heuristic grounds as part of  a model of asymptotic electrodynamics proposed by the author long ago  \cite{her98}.

Finally, we would like to indicate that the usual procedure of cutting off interaction at large distances, and subsequent `adiabatic limit', could not recover the long-range characteristic, which has been discarded earlier.

\setcounter{subsection}{0}
\renewcommand{\thesubsection}{\Alph{subsection}}

\section*{Appendix}

\subsection{Energy content of the state $\w_h$}\label{omega}

Here we briefly discuss the energy content of the state $\w_h=(\W_h,.\,\W_h)$, on which the representation used in this article is based, see the remarks in Section \ref{elm}. For this we refer the reader to the original formulation of the representation in terms of a direct integral Hilbert space, Theorem 6.3 in \cite{her98}. In this language, for the unitary operators $U(x)$ representing translations,  one has
\begin{equation}
 \w_h(U(x))=\int\exp\Big[\tfrac{i}{2}\{V_f,T_xV_f\}
 -\tfrac{1}{2}(\ti{T_xV_f}-\ti{V_f},\ti{T_xV_f}-\ti{V_f})_\reg\Big]\,d\mu_B(f)\,,
\end{equation}
where $\dV_f(\w,l)=f(l)h(\w,l)$, $f\in\Hc_\ir^r$, and $d\mu_B(f)$ is the Gaussian measure on the space of functions $f\in\Hc_\ir^r$, with the covariance operator $B$. The above expression is a special case of the formula given in the proof of the theorem mentioned above. The term with the symplectic form in the argument of the exponent may be written as
\begin{equation}
 \tfrac{i}{2}\{V_f,T_x V_f-V_f\}
 =\tfrac{1}{2}(\ti{V_f},\ti{T_xV_f}-\ti{V_f})_\reg
 -\tfrac{1}{2}(\ti{T_xV_f}-\ti{V_f},\ti{V_f})_\reg\,,
\end{equation}
where, strictly speaking, $\ti{V_f}$ is not in $\Hc_\reg$, but the product extends to this case because of sufficient regularity of the other function in the product. In this way one finds that the argument of the exponent is equal to
\begin{equation}
 \int
 \int_0^\infty|\ti{h}(\w,l)|^2\big(e^{i\w l\cdot x}-1\big)
 \frac{d\w}{\w}\big(-f(l)^2\big)\,d^2l\,.
\end{equation}
Taking into account that $U(x)=\exp[ix\cdot P]$, with $P$ the energy-momentum operator, and expanding the exponent in the integral in $x$, one finds
\begin{equation}
\begin{aligned}
 \w_h(P_a)&= \int\bigg\{\int
 m_0(l)\,l_a\big(-f(l)^2\big)\,d^2l\bigg\}\,d\mu_B(f)\,,  \\
 \w_h(P_aP_b)&=\int\bigg\{
 m_1(l)\,l_al_b\big(-f(l)^2\big)\,d^2l\\
 &+\int m_0(l)l_a(-f(l)^2)\,d^2l\,
 \int m_0(l')l_b(-f(l')^2)\,d^2l'\bigg\}\,d\mu_B(f)\,,
\end{aligned}
\end{equation}
where
\begin{equation}
 m_0(l)=\int_0^\infty|\ti{h}(\w,l)|^2 d\w\,,\quad
 m_1(l)=\int_0^\infty|\ti{h}(\w,l)|^2 \w\,d\w\,.
\end{equation}
Suppose that $\ti{h}$ has the form $\ti{h}(\w,l)=\ti{h}(\w l^0)$, in which case one has
\begin{equation}
 m_0(l)=\frac{m_0}{l^0}\,,\ m_1(l)=\frac{m_1}{(l^0)^2}\,,\
 m_0=\int_0^\infty|\ti{h}(u)|^2du\,,\ m_1=\int_0^\infty|\ti{h}(u)|^2udu\,.
\end{equation}
Then the properties of Gaussian measures give (see, e.g. \cite{dal91})
\begin{equation}
\begin{aligned}
 \w_h(P^0)&=m_0\int\|f\|^2d\mu_B(f)=m_0\Tr B\,,\\
 \w_h((P^0)^2)&=\int(m_1\|f\|^2+m_0^2\|f\|^4)\,d\mu_B(f)\\
 &=m_1\Tr B+m_0^2[(\Tr B)^2+2\Tr B^2]\,,\\
 \w_h\big([P^0-\w_h(P^0)]^2\big)
 &=m_1\Tr B+2m_0^2\Tr B^2\,.
\end{aligned}
\end{equation}
The constants $m_0$ and $m_1$ are strictly positive, but may be made arbitrarily small by an appropriate choice of the function $\ti{h}$. Thus both the energy content, and its fluctuations, of the state $\w_h$ may be arbitrarily small, but never vanish. There is no vacuum in the representation.

\subsection{Commutator function in almost radial gauge}\label{sform}

For $\kappa(x,l)$ defined in \eqref{ar_kappa} we note the identity
\begin{equation}
 \p_a^x\kappa_b(x,l)=\p_b^l\big[l_a\zeta(x\cdot l,l)\big]\,,
\end{equation}
which follows by the use of the relations in Appendix C in \cite{her22}. It then easily follows for $W(s,l)$ defined in \eqref{ar_WK}, that
\begin{equation}
\begin{aligned}
 \p\big[\zeta(s,l)l\cdot W(s,l)\big]
 =&-\int\delta'(s-x\cdot l)x\zeta(x\cdot l,l)l\cdot K(x)\,dx\\
 &+\int\delta(s-x\cdot l)K(x)\cdot \p_x\kappa(x,l)\,dx\,.
\end{aligned}
\end{equation}
The second formula in \eqref{ar_VK} may be now written as
\begin{align}
 \dV_K(s,l)&=\dU_K(s,l)+Y_K(s,l)\,,\\
 U_K(s,l)&=\int\delta(s-x\cdot l)
 \big[ K(x)-\kappa(x,l)l\cdot K(x)\big]\,dx\,,\\
 Y_K(s,l)
 &=\int\delta(s-x\cdot l)K(x)\cdot \p_x\kappa(x,l)\,dx\,.
\end{align}
We admit timelike tails in $K(x)$ of the form $K^\as(x)=x\sigma(x)$, where $\sigma(x)$ is smooth outside the origin, homogeneous of degree $-4$, and with support inside the light cone. We assume, for simplicity, that the difference $K(x)-K^\as(x)$ decays rapidly in timelike infinity. Then with the use of properties of functions $\zeta$ and $\eta$ one shows, that also $K(x)-\kappa(x,l)l\cdot K(x)$ and $K(x)\cdot \p_x\kappa(x,l)$ decay rapidly, which results in fast decay of $U_K(s,l)$ and $Y_K(s,l)$. Therefore, when $\dV_{K_i}(s,l)$, $i=1,2$, are set into the second form of \eqref{elm_sform}, one can transfer derivatives of $U_{K_i}$ to $\sgn(s-\tau)$, which leads to the formula \eqref{ar_Dcom}.

\subsection{Projection operator $P_\ir$}\label{cone}

Let $f(l)$ be a smooth complex vector function on $C_+$, homogeneous of degree $-1$, and such that $l\cdot f(l)=0$. As shown in Appendix of \cite{her98}, there exist smooth functions $\phi$ and $\psi$, homogeneous of degree $0$, unique up to the addition of constants, such that\footnote{In the reference $f$ is real, but the same is true for complex functions.}
\begin{equation}\label{cone_fpp}
 l_af_b(l)-l_bf_a(l)=L_{ab}\phi(l)-{}^*L_{ab}\psi(l)\,,
\end{equation}
and then for two such functions one has
\begin{equation}
 -\int \ov{f_1(l)}\cdot f_2(l)\,d^2l
 =-\int \p\ov{\phi_1}(l)\cdot \p\phi_2(l)\,d^2l
 -\int \p\ov{\psi_1}(l)\cdot \p\psi_2(l)\,d^2l\,.
\end{equation}
Thus, the transformation $f\mapsto \p\phi$, extended by continuity to the whole Hilbert space $\Hc_0$, is the orthogonal projection $P_\ir:\Hc_0\mapsto\Hc_\ir$. We find its explicit form.
\begin{lem}\label{cone_proj}
 Let $f\in \Cc^\infty\subset \Hc_0$, and for the sake of differentiation extend $f(l)$ to a neighborhood of the cone, respecting the defining properties: $l\cdot f(l)=0$ and homogeneity of degree $-1$. Then $P_\ir f=\p\phi$ (equality in the sense of $\Hc_\ir$---up to addition of a term proportional to $l$), where
\begin{equation}\label{cone_pfromf}
\begin{aligned}
 [\phi(l)]&=\Big[-\frac{1}{4\pi}\int\log\Big(\frac{l\cdot l'}{t\cdot l'}\Big)\,\p\cdot f(l')\,d^2l'\Big]\\
 &=\Big[\frac{1}{4\pi}\int\frac{l\cdot f(l')}{l\cdot l'}\,d^2l'\Big]\,,
\end{aligned}
\end{equation}
and where the square bracket denotes an equivalence class with respect to the addition of a constant. Moreover, one has
\begin{equation}\label{cone_Lpf}
 \big[L^n\phi(l)\big]=\Big[-\frac{1}{4\pi}
 \int\log\Big(\frac{l\cdot l'}{t\cdot l'}\Big)\,
 {L'}^n\p\cdot f(l')\,d^2l'\Big]\,.
\end{equation}
\end{lem}
\begin{proof}
We apply $L_c{}^b$ to the lhs of identity \eqref{cone_fpp}. With the extension of $f$ as assumed in the Lemma, we find
\begin{equation}\label{cone_lf}
 L_c{}^b\big(l_af_b(l)-l_bf_a(l)\big)
 =l_al_c\p\cdot f(l)+l_af_c(l)-l_cf_a(l)\,,
\end{equation}
which shows that the restriction of $\p\cdot f(l)$ back to the cone does not depend on the extension in the assumed class. Similarly, we extend $\phi(l)$ and $\psi(l)$ to a neighborhood of the cone, respecting homogeneity of degree $0$, and apply $L_c{}^b$ to the rhs. The use of identities (A1) and (A2) in Appendix of \cite{her98} gives
\begin{equation}\label{cone_lpp}
 L_c{}^b\big(L_{ab}\phi(l)-{}^*L_{ab}\psi(l)\big)
 =l_al_c\p^2\phi(l)+L_{ac}\phi(l)-{}^*L_{ac}\psi(l)\,,
\end{equation}
which again shows that the restriction of $\p^2\phi(l)$ back to the cone does not depend on the extension used. Equating \eqref{cone_lf} and \eqref{cone_lpp}, and using \eqref{cone_fpp}, one finds that
\begin{equation}\label{cone_ffi}
 \p^2\phi(l)=\p\cdot f(l)\,.
\end{equation}
One can now choose a timelike, future pointing unit vector $t$, and use formula (A8) in \cite{her98} to obtain the inverse formula
\begin{equation}
 \phi(l)=-\frac{1}{4\pi}\int\log\Big(\frac{l\cdot l'}{t\cdot l'}\Big)\,
 \p'\cdot f(l')\,d^2l'+\phi_t\,,
\end{equation}
where $\phi_t$ is an arbitrary constant. Integrating $\p$ by parts by \eqref{int_Lint}, one obtains the second formula in \eqref{cone_pfromf}. Finally, to prove \eqref{cone_Lpf} one notes that
\[
 (L_{ab}+L'_{ab})\log\frac{l\cdot l'}{t\cdot l'}=\frac{t_al'_b-t_bl'_a}{t\cdot l'}\,.
\]
Integrating $L'_{ab}$ by parts one obtains the identity for $n=1$. Further steps are similar.
\end{proof}

\subsection{Asymptotics of a Klein--Gordon wave packet}\label{packet}

Standard statements of the stationary phase method, which is the usual means for the asymptotic expansion of the wave packets, assume compact support of the integrand (see \cite{hoer90}, \cite{vai89}). Going beyond this assumption needs taking into account specific form of the phase function. We do this here for the Klein--Gordon wave packets.\footnote{D. Ruelle in his pioneering paper \cite{rue62} makes use of the specific form of the packet, but keeps the compactness assumption.}
\begin{thm}\label{packet_askg}
Let $f(p)$, $p\in H_m$, be a Schwartz function on the mass hyperboloid, with the invariant measure as defined in \eqref{dir_md}, and for $x^2>0$ denote $x=\la u$, $u\in H_1$, $\la\in \mR\setminus \{0\}$.
Then for each $N\geq0$ we have
\begin{equation}\label{packet_as}
\begin{aligned}
 &(2\pi)^{-\frac{3}{2}}\int f(p)e^{-ix\cdot p}d\mu_m(p)\\
 &=\theta(x^2-1)e^{-i(m\la+\sgn(\la)\frac{3\pi}{4})}\Big(\frac{m}{|\la|}\Big)^\frac{3}{2}\,
 \sum_{k=0}^N\Big(\frac{-i}{m\la}\Big)^kL_k(u)f(mu)\\
 &+O\big((1+|x|)^{-N-\frac{5}{2}}\big)\,,
\end{aligned}
\end{equation}
where $L_0=1$, and $L_k(u)$ for each $k\geq1$ is a differential operator in $u$ of order~$2k$, with real polynomial coefficient functions of $u$ of degree  $\leq2k$.
In~particular, the series contribution is absent outside the support of $f(mu)$.
\end{thm}

\begin{proof}
To simplify notation we write the proof in the units, in which $m=1$, and then $p=v$, $v\in H_1$. First we note that
\begin{equation}\label{packet_idex}
 e^{-ix\cdot v}=-i2v^0
 \frac{x^0\vv+v^0\xx}{|x|^2-x^2|v|^2}
 \cdot\frac{\p}{\p\vv}e^{-ix\cdot v}\,.
\end{equation}
Let $a(s)$ be a smooth function on $\mR$, such that $a(s)=1$ for $s\leq\frac{1}{4}$, and $a(s)=0$ for $s\geq\frac{1}{2}$. We denote $f_1(v,x)=f(v)a(|v|^2x^2/|x|^2)$ and note that on the support of this function the denominator in \eqref{packet_idex} is bounded from below by $|x|^2/2$. Therefore, in the integral with $f$ replaced by $f_1$, one can integrate by parts arbitrarily many times, which shows that this part satisfies the bound of the rest in \eqref{packet_as}.

Consider now the function $f(v)-f_1(v,x)$, whose support is contained in $x^2|v|^2\geq|x|^2/4$, which implies that $x$ is timelike in this region, and we specify to $x^0\geq1$: the case $x^0\leq-1$ is easily then derived with the use of conjugation, and the case $|x^0|<1$ is trivial. We write $x=\la u$, $u\in H_1$, $\la>0$, and denote $f_2(v,u)=f(v)b(|v|^2/|u|^2)$, $b(s)=1-a(s)$. On the support of $f_2(v,u)$ we have $|v|\geq\frac{1}{2}|u|$.

Let $\La$ be the Lorentz boost such that  $u=\La t$, which means that in the given reference system the matrix of this transformation is $\begin{pmatrix}u^0&\uu^\intercal\\\uu&u^0\1\end{pmatrix}$.  Consider the integral
\begin{equation}\label{packet_I}
 I=\int e^{-i\la(u\cdot v-1)}f_2(v,u)\,d\mu(v)
 =\int e^{-i\la(w^0-1)}f_2(\La w,u)\,d\mu(w)\,,
\end{equation}
where $v=\La w$, $w\in H_1$, and the Lorentz invariance of the measure $d\mu(v)$ was used. Changing variables $\ww$ to $r=\sqrt{w^0-1}\in[0,\infty)$ and the spherical angles of $\ww$, we can write this integral as
\begin{equation}
 I=\int\limits_0^\infty F(r)2(2+r^2)^\frac{1}{2}r^2\p_r^{2N+5}e_{2N+5}(\la,r)\,dr\,,\quad
 F(r)=\int f_2(\La w,u)\,d\Omega(\ww)\,,
\end{equation}
where $e_k(\la,r)$ is a primitive of order $k$ for $e^{-i\la r^2}$ ($\la>0$) given by\footnote{At this point we follow \cite{vai89}.}
\begin{equation}
 e_k(\la,r)=\frac{(-1)^k}{(k-1)!}e^{-ik\frac{\pi}{4}}
 \int\limits_0^\infty s^{k-1} e^{-i\la r(s)^2}ds\,,\quad
 r(s)=r+e^{-i\frac{\pi}{4}}s\,
\end{equation}
(use $\p_rr(s)=e^{i\frac{\pi}{4}}\p_sr(s)$ and integrate by parts).
It is easy to show that
\begin{equation}
 |e_k(\la,r)|\leq c_k\la^{-k/2}\,,\quad
 e_{2k+1}(\la,0)
 =-\frac{\sqrt{\pi}}{2^{2k+1}k!}e^{-i(2k+1)\frac{\pi}{4}}\la^{-k-\frac{1}{2}}\,.
\end{equation}
Integrating $\p_r$ in $I$ by parts $2N+5$ times, and taking into account that $F^{(2k+1)}(0)=0$, one obtains the asymptotic series of the thesis ending with $k=N+1$, and the rest in the form (with restored integration variables $\ww$)
\begin{equation}
 -\int \p_r^{2N+5}\big(f_2(\La w,u)(2+r^2)^\frac{1}{2}r^2\big)e_{2N+5}(\la,r)
 \frac{d\mu(w)}{\sqrt{w^0+1}(w^0-1)}\,.
\end{equation}
Going back to variables $\vv$, taking into account that $|v|\geq \frac{1}{2}|u|$ on the support of $f_2$, and also noting that
$
  w^0-1=u\cdot v-1\geq 2|\vv-\uu|^2/(v^0+u^0)^2
$,
one can estimate this rest by
\begin{equation}
 \frac{\con}{\la^{N+\frac{5}{2}}}
 \int\limits_{|v|\geq\frac{1}{2}|u|}\frac{d^3\vv}{|\vv-\uu|^2(v^0)^{N+\frac{5}{2}+3}}
 \leq\frac{\con}{(\la u^0)^{N+\frac{5}{2}}}
 =\frac{\con}{|x^0|^{N+\frac{5}{2}}}\,.
\end{equation}
The last term of the series with $k=N+1$ can be adjoined to the rest without changing the form of bound. Finally, for $0<x^2<1$ ($0<\la<1$) and $x^0\geq1$, the terms of the series are bounded by
$\con\la^{-k-\frac{3}{2}}(u^0)^{-N-\frac{5}{2}}\leq\con(x^0)^{-N-\frac{5}{2}}$, which closes the proof.

\end{proof}

\begin{col}\label{packet_ascur}
 Let $f(p,q)$, $p,q\in H_m$, be a Schwartz function on $H_m\times H_m$, and  for $x^2>0$ denote $x=\la u$, $u\in H_1$, $\la\in \mR\setminus \{0\}$.
Then for each $N>0$ we have
\begin{equation}\label{packet_asplus}
\begin{aligned}
 &(2\pi)^{-3}\int f(p,q)e^{-ix\cdot (p+q)}d\mu_m(p)\,d\mu_m(q)\\
 &=\theta(x^2-1)\Big\{i\sgn(\la)e^{-i2m\la}
 \Big(\frac{m}{|\la|}\Big)^3\,f(mu,mu)+O(|x|^{-4})\Big\}\\
 &+O\big((1+|x|)^{-N}\big)\,,
\end{aligned}
\end{equation}
\begin{equation}\label{packet_asminus}
\begin{aligned}
 &(2\pi)^{-3}\int f(p,q)e^{-ix\cdot (p-q)}d\mu_m(p)\,d\mu_m(q)\\
 &=\theta(x^2-1)\Big\{
 \Big(\frac{m}{|\la|}\Big)^3\,f(mu,mu)+O(|x|^{-4})\Big\}
 +O\big((1+|x|)^{-N}\big)\,,
\end{aligned}
\end{equation}
\end{col}
For the proof it is sufficient to apply the operations used in the proof of Theorem \ref{packet_askg} to $p$ and $q$ separately.

\end{document}